\documentclass [11pt]{article}

\usepackage{amssymb,fullpage,graphics}

\newcommand\Z{\mbox{$\mathbb Z$}}

\newcommand\R{\mbox{$\mathbb R$}}

\newcommand\F{\mbox{$\mathbb{F}$}}
\def\B{\{0,1\}}
\def\bfz{{\bf 0}}
\def\bv{{\bf v}}
\def\poly{\mbox{poly}}
\def\cost{{\rm cost}}
\def\opt{{\rm opt}}

\def\epsilon{\varepsilon}
\def\phi{\varphi}

\def\bpp{{\rm BPP}}
\def\npo{{\rm NPO}}

\def\svp{{\rm SVP}}
\def\cvp{{\rm CVP}}

\def\pcp{{\rm PCP}}
\def\np{{\rm NP}}
\def\p{{\rm P}}
\def\zpp{{\rm ZPP}}
\def\bpp{{\rm BPP}}
\def\qrp{{\rm QRP}}
\def\rp{{\rm RP}}
\def\qp{{\rm QP}}

\newtheorem{theorem}{Theorem}

\newtheorem{claim}[theorem]{Claim}

\newtheorem{definition}[theorem]{Definition}
\newenvironment{proof}{\noindent {\sc Proof:}}{$\Box$ \medskip} 

\newtheorem{problem}{Problem}

\def\xor{\oplus}
\def\implies{\Rightarrow}

\title {\bf\Large Inapproximability of Combinatorial Optimization Problems}

\author {{\large\sc Luca Trevisan}\thanks{{\tt luca@cs.berkeley.edu}.
U.C. Berkeley, Computer Science Division. Work supported by NSF grant CCR-9984703,
a Sloan Research Fellowship, and an Okawa Foundation Grant.}}

\date{July 27, 2004}

\begin {document}

\sloppy

\maketitle

\begin{abstract}
We survey results on the hardness of approximating combinatorial
optimization problems.
\end{abstract}

\tableofcontents

\section {Introduction}


Hundreds of interesting and important combinatorial optimization
problems are \np-hard, and so it is unlikely that any of them
can be solved by an efficient exact algorithm. Short of proving $\p=\np$,
when one deals with an
\np-hard problem one can either hope to design an exact algorithm
that runs in polynomial time on ``many'' instances but has exponential
worst-case running time, or to design an efficient algorithm that
finds sub-optimal solutions. In this paper we focus on {\em approximation
algorithms}, that are  algorithms
of the second kind with a provably good worst-case ratio between the
value of the solution found by the algorithm and the true optimum.

Formally, we say that an algorithm is {\em $r$-approximate}
for a minimization problem (respectively, a maximization problem) if, on every input,
the algorithm finds a solution whose cost is at
most $r$ times the optimum (respectively, at least $1/r$ times the optimum). 
The ratio $r$, which is always $\geq 1$,
 is also called the {\em performance ratio} of the algorithm.

For some problems, it is possible to prove that even the design of
an $r$-approximate algorithm with small $r$ is impossible, unless
$\p=\np$. Results of this kind, called {\em inapproximability} results, are
the subject of this survey.

\subsection{A Brief Historical Overview}

The seeming intractability of many combinatorial optimization problems
was observed already in the 1960s, motivating the development of suboptimal
heuristic algorithms and, in particular, the notion of approximation
algorithm as defined above. An early example of analysis of an approximation algorithm
is a paper of Graham on scheduling problems~\cite{G66}. 

The theory of NP-completeness \cite{C71,L73,K72},
and its success in classifying the complexity of optimization problems,
provided added motivation to the study of efficient suboptimal 
algorithms and to the rigourus
analysis of approximation algorithms. A 1973 seminal paper by
Johnson~\cite{J74}  gave the field its current foundations.
Johnson considers the problems Max SAT, Set Cover, Independent Set,
and Coloring.\footnote{The Max SAT problem is defined in Section \ref{sec:maxsat},
the Independent Set problem is defined in Section \ref{sec:vc}, the Coloring
problem is defined in Section \ref{sec:color}, and the Set Cover 
problem is 
defined in Section \ref{sec:cover}.} He notes that there is  a 2-approximate algorithm for Max SAT,
and that this can be improved to $8/7$ for Max E3SAT, the restriction
of the problem to instances in which every clause contains exactly three
literals. He also presents a $(1+\ln k)$-approximate algorithm for Set Cover,
where $k$ is the size of the largest set in the collection, and he
notes that natural heuristics for Coloring and Independent Set
fail to give even a $n^{1-\epsilon}$ approximation for $\epsilon>0$.

Thirty years after Johnson's paper, the design and analysis of approximation
algorithms has become a large and successful research area. A
book edited by Hochbaum \cite{Hoc96} and two more recent
textbooks \cite{ACGKMP99,V01} give an extensive overview of
the field. The book by Ausiello et al.~\cite{ACGKMP99} also includes
a list of known results for hundreds of problems.

As the field progressed and matured, it became apparent that
different combinatorial optimization problems behaved different
from the point of view of approximability. For some problems,
researchers were able to devise polynomial time approximation
schemes (abbreviated PTAS), that is, to devise an $r$-approximate
algorithm for every $r>1$. For other problems, such as Max SAT and 
Vertex Cover,\footnote{We define the Vertex Cover problem in Section~\ref{sec:vc}.}
$r$-approximate algorithms for constant $r$ were known, but no
approximation scheme. Other problems like Set Cover had no
known constant factor approximation algorithm, but were known
to admit an approximation algorithm with a performance ratio
that was slowly growing (logarithmic) in the input length.
Finally, there were problems like Independent Set and Coloring
for which not even a slowly growing performance ratio was known
to be achievable in polynomial time. On the other hand, until 1990, there
was no negative result showing that the existence of approximation
schemes for any of the above problems would imply $\p=\np$ or other
unlikely consequences. Indeed, few inapproximability results were 
known at all. Furthermore, there was a general intuition
that, in order to prove inapproximability for the above mentioned
problems (and others), it was necessary to depart from 
 the standard techniques used to prove the \np-hardness of problems.

To see the limitations of standard \np-hardness proofs,
consider for example the proof that solving optimally the Independent Set
problem is \np-hard. The proof starts from a 3SAT instance $\phi$
and constructs a graph $G$ and an integer $k$ such that if $\phi$
is satisfiable then $G$ has an independent set of size $k$, and
if $\phi$ is not satisfiable then all independent sets in $G$
have size at most $k-1$. It then follows that a polynomial time exact algorithm
for solving Independent Set, together with the reduction, would
imply a polynomial time algorithm for 3SAT, and, from Cook's theorem,
we would have a polynomial time algorithm for every problem in \np.
Consider now the question of the approximability of the Independent Set problem.
Looking at the reduction from 3SAT
(that we describe in Section \ref{sec:vc}) we can see that
if $\phi$ has an assignment that satisfies all the clauses except $c$ ones,
then $G$ has an independent set of size $k-c$ and, given such an assignment, the
independent set is easy to construct. Furthermore,
the instances of 3SAT produced by Cook's theorem are such that
it is easy to find assignments that satisfy all the clauses but
one. In conclusion,  the instances that we get by reducing a generic \np\
problem to Independent Set are very easy to approximate.

To derive an inapproximability result, we need a much stronger reduction.
Suppose we want to prove that no $2$-approximate algorithm exists for
the Independent Set problem assuming $\p \neq \np$. Then a natural proof
strategy would be  to start from 
an \np-complete problem, say, 3SAT, 
and then reduce an instance $\phi$ of 3SAT
to an instance $G_\phi$ of Independent Set, with the property that, for some
$k$, if $\phi$ is satisfiable then
\begin{equation}
OPT_{\rm Independent\ Set} (G_\phi) \geq k
\end{equation}
and if $\phi$ is not satisfiable then 
\begin{equation}
OPT_{\rm Independent\ Set} (G_\phi) < k/2 \ .
\end{equation}
Suppose we have such a reduction, and suppose we also have a
$2$-approximate algorithm for Independent Set; then the algorithm, given $G_\phi$,
will find a solution of cost at least $k/2$ if and only if $\phi$ is satisfiable. 
The approximation
algorithm then gives a polynomial time algorithm for 3SAT,
and so no such algorithm can exist if $\p \neq \np$.

All the \np-completeness proofs for graph problems before 1990, however, can be essentially
described as follows: we start from the computation of
a generic non-deterministic Turing machine, then we encode its
computation as a 3SAT formula, using the construction of Cook's theorem,
and then we reduce 3SAT to the problem of interest (the reduction may be
presented as a sequence of reductions involving several intermediate problems,
but it can always be thought of as a direct reduction from 3SAT)
 by encoding variables
and clauses of the formula as sub-graphs connected in a proper way.
The computation of a Turing machine is very sensitive to small changes,
and it seems impossible to generate an inapproximability gap starting
from a fragile model and applying  ``local'' reductions. The only inapproximability
results that can be proved with such reductions are for problems that
remain \np-hard even restricted to instances where the optimum is a small
constant. For example, in the Metric Min $k$-Center problem it is \np-hard
to decide whether the optimum has cost 1 or 2, and so no algorithm can have
a performance ratio smaller than 2 unless $\p=\np$ \cite{HS85}. Similarly,
in the Coloring problem it is \np-hard to decide wether the optimum has
cost 3 or 4, and so no algorithm has performance ratio smaller than
$4/3$ unless $\p=\np$, and Garey and Johnson \cite{GJ76} show that the
gap can be ``amplified'' to $k$ versus $2k-4$ for constant $k$, ruling
out also algorithms with performance ratio smaller than 2. 
Most interesting problems, however, become trivial when restricted
to inputs where the optimum is a constant.

To prove more general inapproximability results it
 seemed necessary to first find a machine model for \np\ in which
accepting computations would be ``very far'' from rejecting computations.
Before such a model was discovered, 
an important piece of work on inapproximability was due
to Papadimitriou and Yannakakis, who showed that, assuming that Max 3SAT
does not have a PTAS, then several other problems do not have
a PTAS~\cite{PY91}. Berman and Schnitger \cite{BS92} proved
that if Max 2SAT does not have a PTAS then, for some $c>0$, the
Independent Set problem cannot be approximated within a factor $n^c$.

The modern study of inapproximability  was made possible
by the discovery, due to Feige et al.~\cite{FGLSS96} in 1990, that probabilistic
proof systems could give a robust model for \np\ that could be
used to prove an inapproximability result for the Independent Set problem.\footnote{Another,
less influential, connection between probabilistic proof checking and inapproximability
was discovered around the same time by Condon~\cite{C91}.}
A year later, Arora et al.~\cite{AS92,ALMSS92}
proved the PCP Theorem, a very strong characterization of \np\ in terms of proof systems,
and showed how to use the PCP Theorem to prove that Max 3SAT does not have
a PTAS. Using the reductions of \cite{PY91} (and others \cite{PY93:tsp,BP89}),
the PCP Theorem gave inapproximability results for several other problems.

An explosion of applications to other problems soon followed,
such as the work of Lund and Yannakakis \cite{LY94} on Set Cover and Coloring
and the work of Arora et al. \cite{ABSS93} on lattice problems, that appeared in 1993.
Later work focused on stronger inapproximability results, obtained both by
proving stronger versions of the PCP Theorem and by devising better reductions.
Despite great technical difficulties, very rapid progress was made in the mid 1990s.
It is interesting to read a survey paper by Bellare \cite{B96},
written in the middle of the most exciting period of that time. Some of the
open questions raised in the paper were solved a few months later.
We now know inapproximability results for Max 3SAT \cite{H97},
Set Cover \cite{F98}, Independent Set \cite{H96} and Coloring \cite{FK96}
showing that the performance ratios of the algorithms
presented in Johnson's paper \cite{J74} are best possible (assuming
$\p\neq\np$ or similar complexity assumptions).

Despite all these achievements, the study of inapproximability results
is still a vibrant field, with radically new insights and major technical
achievements emerging at a steady pace.

\subsection{Organization of This Survey}

To give justice to the breadth and the diversity of this field, we chose
to  present it from different perspectives.

After a review of technical definitions
(Section \ref{sec:prelim}),
we  disucss the PCP Theorem in Section~\ref{sec:pcp},
inapproximability results that follow from the PCP Theorem in Section~\ref{sec:reductions},
and stronger results that follow
from ``optimized'' versions of the PCP theorem and/or from ``optimized'' reductions
in Section~\ref{sec:reductions-b}.
This is a ``textbook'' presentation of inapproximability results,
similar, for example, to the one in \cite[Chapter 29]{V01}. We include
some proofs and we discuss problems for which the reductions are
relatively simple and yield inapproximability results that are tight or almost tight.
The focus of Sections~\ref{sec:pcp}, \ref{sec:reductions}, and \ref{sec:reductions-b} is
 on {\em techniques} used to prove inapproximability
results.

In Section \ref{sec:list} we review what is known for various fundamental
problems. This section has mostly a reference character, and the focus
is on {\em problems}. We give particular space to problems for which 
there are important open questions and that the reader could use
as research topics.

After approaching the field from the perspectives of techniques and
of results for specific problems, we discuss in Section \ref{sec:misc}
a number of alternate questions
that have been pursued. One question relates
to the fact that the best possible performance ratios for
natural problems that do not have a PTAS were observed 
to often be  $O(1)$, $O(\log n)$ or  $n^{O(1)}$. 
A survey paper by Arora and Lund~\cite{AL96} emphasizes this regularity.
For a subclass of optimization problems it was actually proved~\cite{C95,KSTW00}
that only a finite number of approximability behaviour were possible.
However, we now
know of a natural problem for which the best possible performance ratio
is $O(\log^* n)$, of another for which it  is  somewhere between $O((\log n)^2)$
and $O((\log n)^3$, and another for which it is somewhere
between $O(\log\log n)$ and $O(\log n / \log\log n)$. Contrary to previous
intuition, it now appears that it is not possible to classify the approximability
of natural problems into a small number of simple cases.
Another topic that we discuss is the relation between instances that are
``hard'' for linear and semidefinite relaxation and instances
that are generated by approximation preserving reductions.
We also discuss the study of {\em complexity classes} of combinatorial
optimization problems, of relations between average-case complexity
and inapproximability, and of the issue of {\em witness length} in PCP constructions.

We make some concluding
remarks  and we discuss a small sample of open questions
in section~\ref{sec:conclusions}.

\subsection{Further Reading}

Several excellent surveys on inapproximability results have been written.

A paper by Arora and Lund~\cite{AL96} gives a very good overview
of the techniques used to prove inapproximability results. Our
discussion in Section~\ref{sec:unusual} is motivated by the classification
scheme defined in \cite{AL96}. Arora has written two more
survey papers~\cite{A98:survey,A98:survey} on PCP and inapproximability. 

The subject of strong inapproximability results is discussed in a paper
by Bellare~\cite{B96}, that we have already mentioned, and in a more
recent paper by Feige \cite{F02:icm}.

A book by Ausiello and others \cite{ACGKMP99} has an extensive compendium
of known inapproximability results for hundreds of problems\footnote{At the
time of writing, the compendium is available at
{\sf http://www.nada.kth.se/$\sim$viggo/problemlist/compendium.html}.} and chapters devoted
to PCP, inapproximability results, and complexity classes of optimization
problems. The books by Papadimitriou \cite{P94} and  by Vazirani \cite{V01}  
have chapters devoted to inapproximability results.

\section{Some Technical Preliminaries}
\label{sec:prelim}

We assume the reader is familiar with the basic notions of 
algorithm, running time and big-Oh notation.  Such
background can be found, for example, in the introductory
chapter of Papadimitriou's textbook \cite{P94}.

\subsubsection*{NP-Completeness}

We assume that all combinatorial objects that we refer to (graphs,
boolean formulas, families of sets) are represented as binary strings.
For a binary string $x$, we denote its length as $|x|$.
We represent a decision problem as a language, that is, as the
set of all inputs for which the answer is YES. We define $\p$ as
the class of languages that can be decided in polynomial time.
We define $\np$ as
the class of languages $L$ such that there is a  polynomial time
computable predicate $V$ and a polynomial $q()$ such that $x\in L$ if and only
if there is $w$, $|w| \leq q(|x|)$ such that $V(x,w)$ accepts. We think
of $w$ as a {\em proof}, or {\em witness} that $x$ is in the language. 

For two languages $L_1$ and $L_2$, we say that $L_1$ reduces to $L_2$,
and we write  $L_1 \leq_m L_2$ if there is polynomial
time computable $f$ such that $x\in L_1$ if and only if $f(x) \in L_2$.
A language $A$ is $\np$-hard if every language $L$ in $\np$ reduces to $A$.
A language is $\np$-complete if it is $\np$-hard and it belongs to $\np$.

\subsubsection*{NPO Problems}

A combinatorial {\em optimization problem} $O$ is defined\footnote{Other conventions are
possible, which are typically equivalent to the one we describe here.}
by a cost function $\cost_O()$ that given an {\em instance} $x$ of the
problem and a {\em solution} $s$ outputs $\cost_O(x,s)$ which is either
the cost of the solution (a non-negative real number)
or the special value $\bot$ if $s$ is not a {\em feasible}
solution for $x$. For every $x$, there is only a finite number of feasible solutions $s$
such that $\cost_O(x,s)\neq \bot$.
If $O$ is a {\em maximization} problem (respectively, {\em maximization}), then 
our goal is, given $x$ to find a feasible solution $s$ such that
$\cost(x,s)$ is smallest (respectively, largest). We denote by $\opt_O(x)$
the cost of an optimal solution for $x$.

For example, in the independent set problem, an instance is an undirected
graph $G=(V,E)$, a feasible solution is a subset $S\subseteq V$ such that
no two vertices of $S$ are connected by an edge. The cost of a feasible $S$
is the number of vertices in $S$.

An optimization problem $O$ is an $\np$-optimization problem, and it belongs
to the class $\npo$, if $\cost_O ()$ is computable in polynomial time and if
 there is a polynomial $q$ such that for every instance
and every solution $s$ that is feasible for $x$ we have $|s| \leq q(|x|)$.
The independent set problem is clearly an $\npo$ problem.

If $\p=\np$, then for every $\npo$ problem there is a polynomial time
optimal algorithm. If $\p\neq \np$, then none of the \npo\ problems
considered in this paper has a polynomial time optimal algorithm.

\subsubsection*{Approximation}

If $O$ is an $\npo$ minimization problem, $r>1$, and $A$ is an algorithm,
we say that $A$ is an {\em $r$-approximate} algorithm for $O$ if, for every instance
$x$, $A(x)$ is a feasible solution for $x$ and $\cost_O(x,A(x)) \leq r \cdot \opt_O(x)$.
If $O$ is a maximization problem and $r\geq 1$, then we say that $A$ is an
$r$-approximate algorithm if $\cost_O(x,A(x)) \geq  \opt_O(x)/r$.
Another common convention, that we do not use here, is to say that 
an algorithm is $r$-approximate for a maximization $\npo$ problem $O$,
with $r\leq 1$ if, for every $x$, $\cost_O(x,A(x)) \geq  r \cdot \opt_O(x)$.
Finally, sometimes we let $r=r(n)$ be a function of the length of the instance $x$, or
of some other parameter related to the size of $x$.

For example, a 2-approximate algorithm for the independent set problem would
be an algorithm that given a graph $G$ outputs a feasible independent set $S$
such that $|S| \geq |S^*| /2$, where $S^*$ is an optimal independent set. As we will
se later, such an approximation algorithm can exist if and only if $\p = \np$.

\subsubsection*{Complexity Classes}

We have already defined the classes
\np\ and \p. We also define, for later reference, classes
of problems solvable by efficient probabilistic algorithms. We
define \bpp\ as the class of decision problems (languages) that
can be solved in polynomial time by a probabilistic algorithm that,
on every input, has a probability of error at most 1/3.
We also define two subclasses
of \bpp. The class \rp\ contains decision problems that admit
a polynomial time probabilistic algorithm that is correct with 
probability one if the correct answer is ``NO,'' and is correct
with probability at least 1/2 if the correct answer is ``YES.''
The class \zpp\ contains decision problems that admit a probabilistic
algorithm that, on every input, runs in average polynomial time
and always returns the correct answer. (The average is taken only
over the random choices of the algorithm.)

We say that an algorithm runs in quasi-polynomial time if, for some constant $c$, it
runs in time $O(n^{(\log n)^c})$ on inputs of length $n$.
We denote by $\qp$ the class of decision problems solvable by
deterministic quasi-polynomial time algorithms, and by $\qrp$
the relaxation of $\rp$ to quasi-polynomial running time.

It is considered extremely unlikely that $\np$ be contained in any
of the complexity classes defined above. Furthermore, if, say $\np\subseteq \rp$,
then for every $\npo$ problem there is a quasi-polynomial time probabilistic
algorithm that on every input outputs, with high probability, an optimal solution.

\section{Probabilistically Checkable Proofs}
\label{sec:pcp}

As discussed in the introduction, probabilistically checkable
proofs (PCPs) provide a ``robust'' characterization of the class \np.
When we reduce a generic \np\ problem $L$ to 3SAT using Cook's theorem,
we give a way to transform an instance $x$ into a 3CNF formula $\phi_x$
so that if $x\in L$ then $\phi_x$ is satisfiable, and if $x\not\in L$
then $\phi_x$ is not satisfiable. Following the proof of Cook's theorem,
however, we see that it is always easy (even when $x\not\in L$) 
to construct an assignment
that satisfies all the clauses of $\phi_x$ except one. 

Using the PCP Theorem one can prove a stronger version of Cook's theorem,
that states that, in the above reduction, if $x\in L$ then $\phi_x$
is satisfiable, and if $x\not\in L$ then there is no assignment
that satisfies even a $1-\epsilon$ fraction of clauses of $\phi_x$,
where $\epsilon>0$ is a fixed constant that does not depend on $x$.
This immediately implies that Max 3SAT does not have a PTAS (unless $\p=\np$), and
that several other problems do not have a PTAS either (unless $\p=\np$), using
the reductions of Papadimitrious and Yannakakis \cite{PY91} and others.

We define PCPs by considering  a probabilistic modification of the
definition of \np. We consider probabilistic polynomial time
verifiers  $V$ that are given an 
input $x$ and ``oracle access'' to
a witness string $w$. We model the fact that $V$ is a probabilistic
algorithm by assuming that $V$, besides the input $x$ and the witness $w$,
takes an additional input $R$, that is a sequence of random bits. Then
$V$ performs a deterministic computation based on $x$, $w$ and $R$.
For fixed $w$ and $x$, when we say ``$V^w(x)$ accepts'' we mean ``the event
that $V$ accepts when given oracle access to witness $w$, input $x$,
and a uniformly distributed random input $R$.'' When we refer to the
``probability that $V^w(x)$ accepts,'' we take the probability over the choices of $R$.

We say that a verifier is $(r(n),q(n))$-restricted if, for every
input $x$ of length $n$ and for every $w$, $V^w(x)$ makes
at most $q(n)$ queries into $w$ and uses at most $r(n)$ random bits.

We define the class $\pcp[r(n),q(n)]$ as follows. A language $L$
is in $\pcp[r(n),q(n)]$ if there is an $(r(n),q(n))$-restricted
verifier $V$ such that if $x\in L$, then there is $w$ such that $V^w(x)$
accepts with probability 1 and if $x\not\in L$ then for every $w$
the probability that $V^w(x)$ accepts is $\leq 1/2$.

We also consider the following more refined notation.
We say that a language $L$ is in $\pcp_{c(n),s(n)} [r(n),q(n)]$
if there is an $(r(n),q(n))$-restricted
verifier $V$ such that if $x\in L$, then there is $w$ such that $V^w(x)$
accepts with probability at least $c(n)$, and if $x\not\in L$ then for every $w$
the probability that $V^w(x)$ accepts is at most $s(n)$.
Of course, the definition makes sense only if $0\leq s(n) < c(n) \leq 1$ for
every $n$. The parameter $c(n)$ is called the {\em completeness} of the verifier
and the parameter $s(n)$ is called the {\em soundness error}, or simply
the {\em soundness} of the verifier.

 Note that if $r(n)= O(\log n)$ then the proof-checking can be {\em derandomized},
that is, $V$ can be simulated by a polynomial time deterministic verifier that
simulates the computation of $V$ on each of the $2^{r(n)} = n^{O(1)}$ possible
random inputs and then computes the probability that $V^w(x)$ accepts, and
then accepts if and only if this probability is one. It then follows
that, for example, $\pcp[O(\log n),O(\log n)] \subseteq \np$. The PCP
Theorem shows a surprising converse.

\begin{theorem}[PCP Theorem] 
$\np = \pcp [O(\log n),O(1)]$.
\end{theorem}

The theorem was proved in \cite{AS92,ALMSS92}, motivated by a  relation 
between PCP and approximation discovered in \cite{FGLSS96}. 
The actual proof relies on previous work, as well as on several new results.
A survey paper by David Johnson \cite {J92:pcp} and the 
introductions of the theses of Madhu Sudan \cite{S92} and Sanjeev Arora \cite{A94} give
an overview of the line of work that led to the PCP Theorem.

\subsection{PCP and the Approximability of Max SAT}
\label{sec:maxsat}

In the Max 3SAT problem we are given a 3CNF boolean formula,
that is, a boolean formula in conjunctive normal form
(AND-of-OR of literals, where a literal is either a variable
or the negation of a variable) such that each term in the
conjunction is the OR of at most three literals. The goal
is to find an assignment that satisfies the largest possible
number of terms.

\begin{theorem} \label{thm:3sat}
The \pcp{} Theorem implies that there is an $\epsilon_1 > 0$ such that there is
no polynomial time $(1+\epsilon_1)$-approximate algorithm for MAX-3SAT, unless $\p=\np$.
\end{theorem}

\begin{proof}
Let $L\in \pcp [O(\log n),q]$ be an \np-complete problem, where $q$ is a constant, and let
$V$ be the $(O(\log n),q)$-restricted verifier for $L$.
We describe a reduction
from $L$ to Max 3SAT. 
Given an instance $x$ of $L$, our plan is to construct a 3CNF formula $\phi_x$ with
 $m$ clauses such that, for some $\epsilon_1 > 0$ to be determined,
\begin{equation} \label{eq:gap}
\begin{array}{rcl}
x \in L & \implies & \phi_x \,\,\mbox{is satisfiable} \\
x \notin L & \implies & \,\mbox{no assignment satisfies more than}\,\, (1-\epsilon_1)m \,\,\mbox{clauses of}\,\, \phi_x.
\end{array}
\end{equation}
Once  (\ref{eq:gap}) is proved, the theorem follows.

We enumerate all random inputs $R$ for $V$. The length of each string is $r(|x|) = O(\log |x|)$, so the number of such strings is polynomial in $|x|$. For each $R$, $V$ chooses $q$ positions $i^R_1, \ldots, i^R_q$ and a Boolean function $f_R:\B^q \to \B$ and accepts iff $f_R(w_{i^R_1}, \ldots, w_{i^R_q})=1$.

We want to simulate the possible computation of the verifier (for different random inputs $R$ and different witnesses $w$) as a Boolean formula. We introduce Boolean variables $x_1, \ldots, x_l$, where $l$ is the length of the witness $w$. 

For every $R$ we add clauses that represent the constraint 
$f_R(x_{i^R_1}, \ldots, x_{i^R_q})=1$. 
This can be done with a CNF of size $\leq 2^q$, that is, we would  need to add
at most $2^q$ clauses if we were allowed to write a $q$CNF expression. 
But we also need to ``convert'' clauses of length $q$ to clauses length $3$, which can
be done by introducing additional variables, as in the standard
reduction from $k$SAT to 3SAT (for example $x_2 \lor x_{10} \lor x_{11} \lor x_{12}$ becomes $(x_2 \lor x_{10} \lor y_R) \land (\bar{y}_R \lor x_{11} \lor x_{12})$). Overall, this
transformation creates a formula $\phi_x$ with  at most $q 2^q$ 3CNF clauses. 

Let us now see the relation between the optimum of $\phi_z$ as an instance
of Max 3SAT and the question of whether $z\in L$.

If $z \in L$, then there is a witness $w$ such that $V^w(z)$ accepts for every $R$. 
Set $x_i = w_i$, and set the auxiliary variables appropriately, then
the assignment satisfies  all clauses, and $\phi_z$ is satisfiable.

If $z \notin L$, then consider an arbitrary assignment to the variables
$x_i$ and to the auxiliary variables, and consider the string $w$ where $w_i$
is set equal to $x_i$. The witness $w$ makes the verifier reject for half of the 
$R \in \B^{r(|z|)}$, and  for each such $R$, one of the clauses representing $f_R$ fails. Overall, at least a fraction $\epsilon_1 = \frac{1}{2} \frac{1}{q 2^q}$ of clauses fails.
\end{proof}

Interestingly, the converse also holds: any gap-creating reduction from an \np-complete
problem to Max 3SAT implies that  the \pcp{} Theorem must be true.

\begin{theorem}
If there is a reduction as in (\ref{eq:gap}) for some problem $L$ in \np, then $L \in \pcp[O(\log n), O(1)]$. In particular, if $L$ is \np-complete then the  \pcp{} Theorem holds.
\end{theorem}

\begin{proof}
We describe how to construct a verifier for $L$.
$V$ on input $z$ expects $w$ to be a satisfying assignment for $\phi_z$. $V$ picks $O(\frac{1}{\epsilon_1})$ clauses of $\varphi_z$ at random, and checks that $w$ satisfies all of them. The number of random bits used by the verifier  is $O(\frac{1}{\epsilon_1} \log m) = O(\log |z|)$.
The number of bits of the witness that are read by the verifier
is   $O(\frac{1}{\epsilon_1}) = O(1)$.
\[\begin{array}{ll}
z \in L & \,\implies\, \phi_z \,\mbox{is satisfiable} \nonumber \\
        & \,\implies\, \exists w \,\,\mbox{such that}\, V^w(z) \,\mbox{always accept.} \nonumber \\
\nonumber \\
z \notin L & \,\implies\, \forall w \,\, \mbox{a fraction}\, \epsilon_1 \,\mbox{of clauses of $\phi$ are unsatisfied by $w$} \nonumber \\
           & \,\implies\, \forall w\,\, V^w(z) \,\mbox{ rejects with probability}\, \geq  \frac{1}{2} \quad  \nonumber
\end{array} \]
\end{proof}

\section{Basic Reductions}
\label{sec:reductions}

We have seen that the PCP Theorem is equivalent to the inapproximability of
Max 3SAT and other constraint satisfaction problems. In this section
we will see several reductions that prove inapproximability results
for other problems.

\subsection{Max 3SAT with Bounded Occurrences}

We begin with a reduction from the Max E3SAT problem on general instances
to the restriction of Max E3SAT to instances in which every variable
occurrs only in a bounded number of clauses. The latter problem will
be a useful starting point for other reductions.

For the reduction we will need {\em expander graphs} of the following type.

\begin{definition}[Expander Graph]
An undirected  graph $G=(V,E)$ is a 1-expander if, for every subset $S\subseteq V$,
$|S| \leq |V|/2$, the number of edges $e(S,V- S)$ having one endpoint in $S$ and
one in $V-S$ is at least $|S|$. 
\end{definition}
\noindent For our purposes, it will be acceptable for the expander graph to have
multiple edges. It is easy to prove the existence of constant-degree 1-expanders
using the probabilistic method. Polynomial-time constructible 1-expanders of
constant degree can be derived from~\cite{GG81}, and, with a smaller degree,
from~\cite{LPS88}. Let $d$ be a constant for which degree-$d$ 1-expanders
can be constructed in polynomial time. ($d=14$ suffices using the
construction of \cite{LPS88}.)

Let now $\phi$ be an instance of 3SAT
with $n$ variables $x_1,\ldots,x_n$ and $m$ clauses.
For each variable $x_i$, let $occ_i$ be the number
of {\em occurrences} of $x_i$, that is, the number of clauses
that involve the literal $x_i$ or the literal $\bar x_i$. 
We write $x_i \in C_j$ if the variable $x_i$ occurs in clause $C_j$.
Notice that $\sum_{i=1}^n occ_i = 3m$.
For each $i$, construct a 1-expander graph $G_i = (V_i,E_i)$ where
$V_i$ has $occ_i$ vertices, one for each occurrence of $x_i$ in $\phi$.
We denote the vertices of $V_i$ as pairs $[i,j]$ such that $x_i$
occurrs in $C_j$. Each of
these graphs has constant degree $d$.

We define a new instance $\psi$ of Max E3SAT
with $N=3m$ variables $Y=\{y_{i,j}\}_{i\in [n],x_i \in C_j}$, one for
each occurrence of each variable in $\phi$. 
For each clause of $\phi$ we put an equivalent
clause in $\psi$. That is, if $C_j = (x_a \vee x_b \vee x_c)$ is a
clause in $\phi$, then $(y_{a,j} \vee y_{b,j} \vee y_{c,j})$ is
a clause in $\psi$. We call these clauses the {\em primary clauses} of $\psi$.
Note that each variable of $\psi$ occurs only in one
primary clause. 

To complete the construction of $\psi$, for every variable $x_i$ in $\phi$, and for every edge
$([i,j],[i,j'])$ in the graph $G_i$, we add the clauses
$(y_{i,j} \vee \bar y_{i',j})$ and $(\bar y_{i,j} \vee y_{i',j})$
to $\psi$. We call these clauses the {\em consistency clauses} of $\psi$.
Notice that if $y_{i,j} = y_{i',j}$ then both consistency clauses are
satisfied, while if $y_{i,j} \neq y_{i',j}$ then one of the two consistency
clauses is contradicted.

This completes the construction of $\psi$. By construction, every variable occurrs in 
at most $2d+1$ clauses of $\psi$, and $\psi$ has $M = m + 3dm$ clauses.

We now claim that the cost of an optimum solution in $\psi$ is determined
by the cost of an optimum solution in $\phi$ and, furthermore, that a 
good approximation algorithm applied to $\psi$ returns a good approximation
for $\phi$. We prove the claim in two steps.

\begin{claim}
If there is an assignment for $\phi$ that satisfies $m-k$ clauses, 
then there is an assignment for $\psi$ that satisfies
$\geq M-k$ clauses.
\end{claim}

\begin{proof} This part of the proof is simple: take the assignment
for $\phi$ and then for every variable $y_{i,j}$ of $\psi$ give to it
the value that the assignment gives to $x_i$. This assignment satisfies
all the consistency clauses and all but $k$ of the remaining clauses.
\end{proof}

\begin{claim}
If there is an assignment for $\psi$ that leaves $k$ clauses
not satisfied, then there is an assignment for $\phi$ that leaves
$\leq k$ clauses not satisfied.
\end{claim}

\begin{proof}
This is the interesting part of the proof. Let $a_{i,j}$ be the
value assigned to $y_{i,j}$. We first ``round'' the assignment
so that all the consistency clauses are satisfied. This is done by
defining an assignment $b_i$, where, for every $i$,
the value $b_i$ is taken to be the majority value
of $a_{i,j}$ over all $j$ such that $x_i\in C_j$, and we assign the value $b_i$ to all
the variables $y_{i,j}$. The assignment $b_i$ satisfies all the
consistency clauses, but it is possible that it contradicts
some primary clauses that were satisfied by $a_{i,j}$.
We claim that, overall, the $b_i$ assignment satisfies
at least as many clauses as the $a_{i,j}$ assignment.
Indeed, for each $i$, if $b_i$ differs from the $a_{i,j}$
for, say, $t$ values of $j$, then there can be at most $t$
primary clauses that were satisfied by $a_{i,j}$ but
are contradicted by $b_i$. On the other hand, 
because of the consistency clauses being laid out as the
edges of a 1-expander graph, at least $t$ consistency
clauses are contradicted by the $a_{i,j}$ assignment
for that value of $i$ alone, and so, the $b_i$ assignment
can be no worse. 

We conclude that $b_i$ assignment contradicts no more
clauses of $\psi$ than are contradicted by $a_{i,j}$,
that is, no more than $k$ clauses. When we apply $b_i$
as an assignment for $\phi$, we see that $b_i$ contradicts
 at most $k$ clauses of $\phi$.
\end{proof}

In conclusion:

\begin{itemize}
\item If $\phi$ is satisfiable then $\psi$ is satisfiable;
\item If every assignment contradicts at least an $\epsilon$ fraction
of the clauses of $\phi$, then every assignment contradicts at
least an $\epsilon/(1+3d)$ fraction of the clauses of $\psi$.
\end{itemize}

\begin{theorem}\label{th:satd}
There are constants $d$ and $\epsilon_2$ and a polynomial time computable
reduction from 3SAT to Max 3SAT-$d$ such that if $\phi$ is satisfiable
then $f(\phi)$ is satisfiable, and if $\phi$ is not satisfiable
then the optimum of $f(\phi)$ is less than $1-\epsilon_2$ times the
number of clauses. In particular, if there is an approximation algorithm
for Max 3SAT-$d$ with performance ratio better than $(1-\epsilon_2)$, then
$\p=\np$.
\end{theorem}

\subsection{Vertex Cover and Independent Set}
\label{sec:vc}

In an undirected graph $G=(V,E)$ a vertex cover
is a set $C\subseteq V$ such that for every edge $(u,v)\in E$
we have either $u\in C$ or $v\in C$, possibly both.
An independent set is a set $S\subseteq V$ such that
for every two vertices $u,v\in S$ we have $(u,v)\not\in E$.
It is easy to see that a set $C$ is a vertex cover
in $G$ if and only if $V-C$ is an independent set.
It then follows that the problem of finding a minimum
size vertex cover is the same as the problem of finding
a maximum size independent set. From the point of view of
approximation, however, the two problems are not equivalent:
the Vertex Cover problem has a 2-approximate algorithm
(but, as we see below, it has no PTAS unless $\p=\np$), while the
Independent Set problem has no constant-factor
approximation unless $\p=\np$.

We give a reduction from Max E3SAT to Independent Set.
The reduction will also prove intractability of
Vertex Cover. If we start from
an instance of Max E3SAT-$d$ we will get a bounded degree
graph, but the reduction works in any case.
The reduction appeared in \cite{PY91}, and it is similar
to the original proof of \np-completeness of
Vertex Cover and Independent Set~\cite{K72}.

Starting from an instance $\phi$ of E3SAT  with $n$ variables
and $m$ clauses, we constuct a graph with $3m$ vertices;
the graph has a vertex $v_{i,j}$ for every occurrence
of a variable $x_i$ in a clause $C_j$. For each clause
$C_j$, the three vertices corresponding to the three
literals in the clause are joined by edges, and form
a triangle (we call such edges {\em clause edges}). 
Furthermore, if a variable $x_i$ occurrs
positively in a clause $C_j$ and negated in a clause
$C_{j'}$, then there is an edge between
the vertices $v_{i,j}$ and $v_{i,j'}$ (we call such 
edges {\em consistency edges}). Let us
call this graph $G_\phi$. 
See Figure~\ref{fig:construction} for an example
of this construction.

{\begin{figure} 
  \centerline{\vbox to 1.5in {\vfil \includegraphics{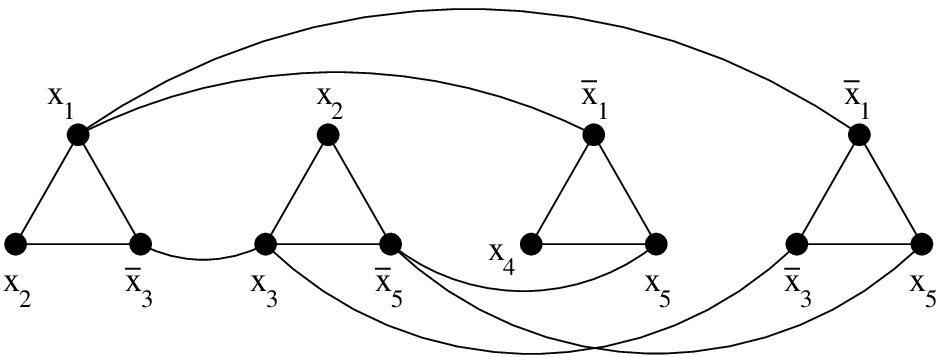} }} 
\caption{Graph construction corresponding to the 3CNF formula $\varphi = (x_1~\lor~x_2~\lor~\bar{x}_3) \land (x_2~\lor~x_3~\lor~\bar{x}_5) \land (\bar{x}_1~\lor~x_4~\lor~x_5) \land (\bar{x}_1~\lor~\bar{x}_3~\lor~x_5)$.}
\label{fig:construction} 
  \end{figure}}

Note that
if every variable occurrs in at most $d$ clauses
then the graph has degree at most $d+2$.

\begin{claim}
There is an independent set of size $\geq t$ in $G_\phi$
if and only if there is an
assignment that satisfies $\geq t$ clauses in $\phi$.
\end{claim}

\begin{proof}
Suppose we have an assignment $a_i$ that satisfies $t$ clauses. For each
clause $C_j$, let us pick a vertex $v_{i,j}$ that corresponds to
a literal of $C_j$ satisfied by $a_i$. We claim that the set of picked
vertices is an independent set in $G_\phi$. To prove the claim, we note
that we picked at most one vertex from each triangle, so that we do not
violate any clause edge, and we picked vertices consistent with the assignment,
so that we could not violate any consistency edge.

For the other direction, suppose we have an independent set with $t$ vertices.
The vertices must come from $t$ different triangles, corresponding to $t$
different clauses. We claim that we can satisfy all such clauses.
We do so by setting an assingment so that $x_i$ takes a value consistent
with the vertices $v_{i,j}$ in the independent set, if any. Since consistency
edges cannot be violated, this is a well defined assignment, and it satisfies
$t$ clauses.
\end{proof}

If we combine this reduction with Theorem \ref{th:satd}, we get the
following result.

\begin{theorem}\label{th:vc}
There is a polynomial time computable
function  mapping instances $\phi$ of 3SAT into graphs $G_\phi$ of maximum degree $d+2$
such that if $\phi$ is satisfiable
then $G_\phi$ has an independent set of size at least $N/3$ (and a vertex
over of size at most $2N/3$, where $N$
is the number of vertices,
 and if $\phi$ is not satisfiable
then every independent set in $G_\phi$ has size at most  $N\cdot (1-\epsilon_2)/3$,
and every vertex cover has size at least $N\cdot (2 + \epsilon_2)/3$. In particular, if there is an approximation algorithm
for Independent Set in degree-$(d+2)$ graphs with performance ratio 
better than $1/(1-\epsilon_2)$, or if there is an approximation
algorithm for Vertex Cover in degree-$(d+2)$  graphs with performance
ratio better than $1+\epsilon_2/2$, then $\p=\np$.
\end{theorem}

\subsection{Steiner Tree}

In the Steiner tree problem we are given a graph $G=(V,E)$,
with weights on the edges, and a subset $C\subseteq V$
of vertices. We want to find the tree of minimum cost
that contains all the vertices of $C$. This problem is 
different from the minimum spanning tree problem because
the tree is not required to contain the vertices in $V-C$,
although it may contain some of them if this is convenient.
An interesting special cases (called {\em Metric Steiner Tree})
arises when $E$ is the complete
graph and the weights are a metric. (That is, they satisfy
the triangle inequality.) Without this restriction, it is
not hard to show that the problem cannot be approximated within any
constant.

We describe a reduction from the Vertex Cover problem in
bounded degree graphs to the Steiner Tree problem.
The reduction  is due to Bern and Plassmann~\cite{BP89}.

We start from a graph $G=(V,E)$, and we assume that $G$
is connected.\footnote{We proved inapproximability of Vertex Cover
without guaranteeing a connected graph. Clearly, if we have
an approximation algorithm that works only in connected graphs,
we can make it work on general graphs with same factor. It follows
that any inapproximability for general graphs implies inapproximability
of connected graphs with the same factor.} 
We define
a new graph $G'$ that has $|V|+|E|$ vertices,
that is, a vertex $[v]$ for each vertex $v$ of $G$ and a
vertex $[u,v]$ for each edge $(u,v)$ of $G$.

The distances in $G'$ are defined as follows:
\begin{itemize}
\item For every edge $(u,v)\in E$, the vertices $[u]$ and $[u,v]$
are at distance one, and so are the vertices $[v]$ and $[u,v]$.

\item Any two vertices $[u]$, $[v]$ are at distance one.

\item All other pairs of vertices are at distance two.

\end{itemize}

We let $C$ be the set of vertices $\{ [u,v] : (u,v)\in E\}$.
This completes the description of the instance of the Steiner Tree problem.
Notice that, since all the distances are either one or two, they
satisfy the triangle inequality, and so the reduction always produces
an instance of Metric Steiner Tree.

\begin{claim}
If there is a vertex cover in $G$ with $k$ vertices,
then there is a Steiner tree in $G'$ of cost $m+k-1$.
\end{claim}

\begin{proof}
Let $S$ be the vertex cover.
Consider the vertices $\{ [v] : v\in S\}$ and the vertices
$\{ [u,v] : (u,v)\in E\}$, and consider the weight-one edges between them.
We have described a connected sub-graph of $G'$, because  every vertex 
in $\{ [u] : u\in S\}$ is  connected to every other vertex in the same set, and every
vertex $[u,v]$ is connected to a vertex in $\{ [u]: u\in S\}$.
Let us take any spanning tree of this subgraph. It has $m+k-1$ edges of weight
one, and so it is of cost $m+k-1$, and it is a feasible solution to the 
Steiner Tree problem.
\end{proof}

\begin{claim}
If there is a Steiner tree in $G'$ of cost $\leq m+k$,
then there is a vertex cover in $G$ with $k$ vertices.
\end{claim}

\begin{proof}
Let $T$ be a feasible Steiner tree. We first modify the tree so that it has no
edge of cost 2. We repeatedly apply the following steps.
\begin{itemize}
\item If there is an edge of cost 2 between a vertex $[w]$
and a vertex $[u,v]$, we remove it and add the two edges
$([w],[u])$ and $([u],[u,v])$ of cost 1.

\item If there is an edge of cost 2 between
a vertex $[u,v]$ and a vertex $[v,w]$,
we remove it and add the two edges $([u,v],[v])$
and $([v],[v,w])$ of cost 1.

\item Finally, and this case is more interesting,
if there is an edge of cost 2 between the vertices
$[u,v]$ and $[w,z]$, we remove the edge, and then
we look at the two connected components into which
$T$ has been broken. Some vertices $[u,v]$ are in
one component, and some vertices are in the other.
This corresponds to a partition of the edges of $G$
into two subsets. Since $G$ is connected, we see
that there must be two edges on different sides
of the partition that share an endpoint. Let
these edges be $(u,v)$ and $(v,w)$ then we can reconnect
$T$ by adding the edges $([u,v],[v])$ and $([v],[v,w])$.

\end{itemize}

We repeat the above steps until no edges of cost two remain.
This process will not
increase the cost, and will return a connected graph. 
We can obtain a tree by removing edges, and improving
the cost, if necessary.

The final tree has only edges of weight one, and it has
a cost $\leq m+k-1$, so it follows that it spans $\leq m+k$
vertices. The $m$ vertices $\{[u,v] : (u,v)\in E\}$
must be in the tree, so the tree has $\leq k$ vertices
$[v]$. Let $S$ be the set of such vertices. We claim that
this is a vertex cover for $G$. Indeed, for every edge
$(u,v)$ in $G$, the vertex $[u,v]$ is connected to $v_0$
in the tree using only edges of weight one, which means
that either $[u]$ or $[v]$ is in the tree, and
that either $u$ or $v$ is in $S$.
\end{proof}

If we combine the reduction with the results of Theorem \ref{th:vc},
we prove the following theorem.

\begin{theorem}
There is a constant $\epsilon_3$ such that if there is a polynomial
time $(1+\epsilon_3)$-approximate algorithm for Metric Steiner
Tree then $\p=\np$.
\end{theorem}

\subsection{More About Independent Set}

In this section we describe a direct reduction from PCP to the Independent
Set problem. This reduction is due to Feige et al.~\cite{FGLSS96}.

Let $L$ be  \np-complete, and $V$ be a verifier showing that
$L\in \pcp_{c,s}[q(n),r(n)]$. For an input $x$, let us consider
all possible computations of $V^w (x)$ over all possible
proofs $w$; a complete description of a computation of $V$ is
given by a specification of the randomness used by $V$, the list
of queries made by $V$ into the proof, and the list of answers.
Indeed,  for a fixed input $x$, each query is determined
by $x$, the randomness, and the previous answers, so that it is
enough to specify the randomness and the answers in order to completely
specify a computation. We call such a description a {\em configuration}.
Note that the total number of configuration is at most $2^{r(n)}\cdot 2^{q(n)}$,
where $n$ is the length of $x$.

Consider now the graph $G_x$ that has a vertex for each {\em accepting}
configuration of $V$ on input $x$, and has an edge between two configurations
$c,c'$ if $c$ and $c'$ are {\em inconsistent}, that is, if $c$ and $c'$ specify
a query to the same location and two different answers to that query. 
We make the following claims.

\begin{claim}
If $x\in L$, then $G_x$ has an independent set of size $\geq c\cdot 2^{r(n)}$.
\end{claim}

\begin{proof}
If $x\in L$, then there is  a proof $w$ such that $V^w(x)$ accepts
with probability at least $c$, that is, there is a proof $w$ such 
that there are at least $c\cdot 2^{r(n)}$ random inputs that
make $V^w(x)$ accept. This implies that there are at least $c\cdot 2^{r(n)}$
mutually consistent configurations in the graph $G_x$, and they
form an independent set.
\end{proof}

\begin{claim}
If $x\not\in L$, then every independent set of $G_x$ has size $\leq s\cdot 2^{r(n)}$.
\end{claim}

\begin{proof}
We prove the contrapositive: we assume
that there is
an  independent set in $G_x$ of size $\geq s\cdot 2^{r(n)}$, and we show that
this implies $x\in L$.
Define a witness $w$ as follows: for every configuration in the independent set,
fix the bits in $w$ queried in the configuration according to the answers in the
configurations. Set the bits of $w$ not queried in any configuration in the
independent set arbitrarily, for example set them all to zero. The $s\cdot 2^{r(n)}$
configurations in the independent set correspond to as many different random strings.
When $V^w(x)$ picks any such   random string, it accepts, and so $V^w(x)$ accepts
with probability at least  $s$, implying $x\in L$.
\end{proof}

It follows that if there is a $\rho$-approximate algorithm for the independent
set problem, then every problem in $\pcp_{c,s} [r(n),q(n)]$ can be
solved in time $\poly( n, 2^{r(n)+q(n)})$, provided $c/s < \rho$.

From the PCP Theorem we immediately get that there cannot be a $\rho$-approximate
algorithm for the independent set problem with $\rho<2$ unless $\p = \np$, but
we can do better.

Suppose that $V$ is a $(O(\log n),O(1))$-restricted verifier for an \np-complete
problem, and that $V$ has soundness $1/2$ and completeness $1$. Define a new
verifier $V'$ that performs two independent repetitions of the computation of $V$,
and that accepts if and only if both repetitions accept. Then $V'$ has clearly
soundness $1/4$ and completeness $1$, and it is still $(O(\log n),O(1))$-restricted,
thus showing that even an approximation better than 4 is infeasible. If we repeat
$V$ a constant number of times, rather than twice, we can rule out any constant
factor approximation for the independent set problem. 

In general, a verifier that makes $k(n)$ repetitions shows that $L\in \pcp_{1,1/2^{k(n)}}
[O(k(n)\cdot \log n), O(k(n))]$, and the reduction to Independent Set produces
graphs that have $2^{O(k(n)\cdot \log n)}$ vertices and for which
$2^{k(n)}$-approximate algorithms are infeasible. If we let $k(n)=\log n$,
then the graph has size $N=2^{O((\log n)^2)}$ and the infeasible ratio is $n$,
which is $2^{\Omega(\sqrt{\log N})}$. So, if we have an algorithm
that on graphs with $N$ vertices runs in polynomial time and has
an approximation ratio $2^{o(\sqrt{\log N})}$, then we have
an $O(n^{O(\log n)})$ algorithm to solve 3SAT, and $\np\subseteq \qp$.
More generally, by setting $k(n)=(\log n)^{O(1)}$, we can show
that if there is an $\epsilon>0$ such that Independent Set can be
approximated within a factor $2^{O((\log n)^{1-\epsilon})}$ then
$\np\subseteq \qp$.

Finally, using random walks in expander graphs as in \cite{IZ89},
it is possible to use the \pcp\ theorem to show that, for every
$k(n)$, $\np = \pcp_{1,1/2^{k(n)}} [O(k(n) + \log n),O(k(n))]$.
If we choose $k(n)=\log n$, then, in the reduction, we have a graph
of size $2^{O(k(n) + \log n)} = n^{O(1)}$ for which an approximation
ratio of $n$ is infeasible. This shows the following result.

\begin{theorem}
There is a constant $c>1$ such that
if there is a polynomial time $n^c$-approximate algorithm for Independent
Set then $\p=\np$.
\end{theorem}

\section{Optimized Reductions and PCP Constructions}
\label{sec:reductions-b}

In this section  we give an overview of tighter inapproximability results
proved with a combination of optimized PCP constructions and optimized
reductions.

\subsection{PCPs Optimized for Max SAT and Max CUT}

In the reduction from a $(O(\log n),q)$-restricted
verifier to Max SAT we lost a factor that
was exponential in $q$. A more careful analysis shows that
the loss depends on the size of the CNF that expresses
a computation of the verifier. To prove stronger inapproximability
results for Max SAT one needs a verifier that, at the same time,
has a good separation between soundness and completeness and
a simple acceptance predicate.  Progress along
these lines is reported in \cite{BGLR93,FK94,BS94,BGS95}.

The optimal inapproximability result for Max 3SAT was proved
by H\aa stad~\cite{H97}, based on the following PCP construction.

\begin{theorem}[H\aa stad \cite{H97}]\label{th:haastad}
 For every $\epsilon >0$, $\np = \pcp_{1-\epsilon,
1/2+\epsilon} [O(\log n),3]$. Furthermore, the verifier behaves
as follows: it uses its randomness to pick three entries $i,j,k$
in the witness and a bit $b$, and it accepts if and only
if $w_i \xor w_j \xor w_k = b$.
\end{theorem}

The furthermore part immediately implies that for
every problem in \np, say, SAT for concreteness, and
for every $\epsilon>0$ there is a reduction that given
a 3CNF formula $\phi$ constructs a system of linear
equations over $GF(2)$ with three variables per equation.
If $\phi$ is satisfiable then there is an assignment to the
variables that satisfies a $1-\epsilon$ fraction of the
equations, and if $\phi$ is not satisfiable then there
is no assignment that satisfies more than a $1/2+\epsilon$
fraction of equations. In particular, it is not possible
to approximate the Max E3LIN-2 problem to within a factor
better than 2 unless $\p=\np$.

To reduce  an instance $I$ of Max E3LIN-2 to an instance $\phi_I$ of
Max 3SAT, we simply take every equation $x_i \xor x_j \xor x_k = b$ in $I$
and express it as  the conjunction of 4 clauses, then $\phi_I$ is the conjunction
of all these clauses. Let $m$ be the number of equations in $I$, then
$\phi_I$ has $4m$ clauses. 
If $\geq m(1-\epsilon)$ of the E3LIN equations
could be satisfied, then $\geq 4m(1-\epsilon)$ of the clauses of $\phi_I$ can
be satsfied, using the same assignment. Conversely,
if it is impossible to satisfy more than
$m(1/2+\epsilon)$ equations in the E3LIN instance $I$  then 
it is impossible to satisfy more than $3.5 m + \epsilon m$ clauses in $\phi_I$.
An approximation algorithm for Max 3SAT with a performance ratio
better than $8/7$ allows to distinguish between the two cases, if $\epsilon$ is small enough,
and so such an algorithm is impossible unless $\p=\np$.

\begin{theorem}
If there is an $r$-approximate algorithm for Max 3SAT, where $r<8/7$, then $\p=\np$.
\end{theorem}

A similar ``local'' reduction is possible from Max E3LIN-2 to Max CUT.
For the purpose of the reduction it is easier to think of Max CUT
as the boolean constraint satisfaction problem where we have 
Boolean variables $x_1,\ldots,x_n$ (one for every vertex in the graph)
and constraints of the form  $x_i \neq x_j$ (one for every edge $(i,j)$ in the
graph); we call constaints of this form ``cut constraints.'' 
To reduce a Max E3LIN-2 instance $I$ to a Max CUT instance $G_I$, we take each 
constraint $x_i \xor x_j \xor x_k = b$ of $I$ and we construct a collection
of cut constraints over $x_i,x_j,x_k$, plus auxiliary variables. The construction
(from \cite{TSSW96})
is such that an assignment to $x_i,x_j,x_k$ that satisfies the E3LIN-2 constraint
with $b=0$ can be extended to the auxiliary variable so that it satisfies 9 of the cut constraints,
while an assignment that contradicts the E3LIN-2 constraint, no matter how it is extended
to the auxiliary variables, satisfies at most 8 cut constraints (and 8 constraints
can always be satisfied by setting the auxiliary variables appropriately). 
When $b=1$, a similar construction exists and the number of satisfied constraints
is 8 versus 7 (instead of 9 versus 8).
In E3LIN-2 instances that are derived from H\aa stad's PCP there are as many
equations with right-hand side 0 as equations with right-hand side 1.
Given all these facts, simple computations show that

\begin{theorem}
If there is an $r$-approximate algorithm for Max CUT, where $r<17/16$,
then $\p=\np$.
\end{theorem}

The best known approximation algorithm for Max CUT is due
to Goemans and Williamson \cite{GW94}, and its performance ratio
is $1/\alpha$ where $\alpha \approx .878$. Khot et al.~\cite{KKMO04}
describe a PCP system that may prove that the algorithm of Goemans 
and Williamson is best possible. In particular, if the ``unique game
conjecture'' is true\footnote{The unique game conjecture, posed
by Khot \cite{K02:unique}, is about the existence of a PCP system for \np\
with certain properties. See also Section \ref{sec:uniquegame}.}
and if a conjecture about extremal properties of monotone functions
is true, then there is no polynomial time approximation algorithm
for Max CUT with performance ratio better than $1/\alpha$, unless $\p=\np$.

\subsection{PCPs Optimized for  Independent Set}
\label{sec:is}

The graph in the reduction of Feige et al. \cite{FGLSS96} has a vertex for every
{\em accepting} computation. This implies that the important parameter
to optimize (in order to derive strong inapproximability results for
Independent Set)  is not the number
of queries but rather the number of accepting computations. This
intuition motivates the following definition.

\begin{definition}[Free Bit Complexity]
We say that a  $(r(n),q(n))$-restricted verifier $V$ for a language $L$
{\em has free bit complexity at most $f(n)$} if, for every input $x$ and every choice of
the random input, there are at most $2^{f(n)}$ possible answers to the
$q(n)$ queries made by the verifier that would make it accept.
\end{definition}

In the reduction from PCP to Independent Set, if we start from
a $(r(n),q(n))$-restricted verifier of free bit complexity $f(n)$,
then we construct a graph with $\leq 2^{r(n)+f(n)}$ vertices,
so that, all other things being equal, a lower free bit
complexity yields a smaller graph and so a stronger relation
between inapproximability and graph size. A further inspection
of the reduction of \cite{FGLSS96} shows that what really
matters is the ratio $f(n)/\log_2 (1/s(n))$, where $s(n)$
is the soundness of the verifier, assuming that the verifier
has completeness 1. This measure is called the {\em amortized
free bits complexity} of the verifier.

\begin{theorem}[Optimized FGLSS Reduction \cite{BGS95}]
Suppose that an  $\np$-complete problem has an $(O(\log n),O(\log n))$-restricted
verifier of amortized free bit complexity $\bar f$, and suppose that, for some
$\epsilon>0$, there is an $n^{\epsilon + 1/(\bar f +1)}$-approximate
algorithm for Independent Set. Then $\np=\zpp$.
\end{theorem}

In order to prove strong inapproximability results it is then enough
to provide a version of the PCP Theorem with a verifier having low amortized
free bit complexity. H\aa stad proved the stronger possible result,
stating that an arbitrarily small amortized free bit complexity is sufficient.

\begin{theorem}[H\aa stad \cite{H96}]
For every $\delta>0$, 3SAT has an $(O(\log n),O(1))$-restricted
verifier of amortized free bit complexity $\leq \delta$.
Therefore, assuming $\zpp\neq \np$, for every $\epsilon>0$
there is no $n^{1-\epsilon}$-approximate algorithm for the
Independent Set problem.
\end{theorem}

The proof of this result was substantially simplified in \cite{ST00,HW01}.

If a graph $G=(V,E)$ has maximum degree $d$, then a maximal independent set
contains at least $|V|/(d+1)$ vertices, and so is a $(d+1)$-approximate
solution. This can be slightly improved, and an $O(d \log\log d /\log d)$-approximate 
algorithm is also known \cite{KMS98,AK98,V96,Hal98}. Trevisan \cite{T01} proves
that no $(d/2^{O(\sqrt {\log d})})$-approximate algorithm
exists unless $\p=\np$.

\section{An Overview of Known Inapproximability Results}
\label{sec:list}

After the overview of techniques given in the previous sections,
we devote this section to an overview of known inapproximability
results for a few selected problems. Some more inapproximability
results for specific problems are also discussed in Section~\ref{sec:unusual}
below.

\subsection{Lattice Problems}

Let $B = \{ \bv_1,\ldots,\bv_n\}$ be a set of linearly
independent vectors in $\R^n$. The {\em lattice} generated
by $B$ is the set of all points of the form 
$\sum_i a_i \bv_i$ where $a_i$ are integers.
For example, suppose $n=2$ and $B = \{ (1/2,0), (0,1/2) \}$. Then
the lattice generated by $B$ contains all points in $\R^2$
whose coordinates are half-integral. 

\subsubsection*{Shortest Vector Problem}

The {\em Shortest Vector Problem}, abbreviated SVP, is the problem of  finding
the shortest non-zero vector of a lattice, given a base for the
lattice. By ``shortest vector,'' we mean the vector of smallest $\ell_2$
norm, where the $\ell_2$ norm of a vector $\bv = (v_1,\ldots,v_n)$ is $|| \bv||_2 = \sqrt{\sum_i v_i^2}$.

SVP is a non-trivial problem because the shortest vector may not be any
of the basis vectors, and may in fact be a complicated linear combination of the basis vectors.
For example, suppose that the basis is $\{ (3,4), (4,5) \}$. Then one
can immediately see that a  shorter vector $(1,1)$ can be derived by
subtracting one base vector from the other; it takes some more attention
to see that a shortest vector can be derived as $(0,1) = 4\cdot (3,4) - 3\cdot(4,5)$.
Of course the complexity of the problem escalates with the number of dimension
and, perhaps surprisingly, the best known approximation algorithms for SVP
achieve only an {\em exponential} approximation as a function of $n$, and
even achieving such an approximation is highly non-trivial \cite{LLL82,S87,AKS01}.
The best known approximation algorithm has performance ratio
  $2^{O(n(\log\log n)/\log n)}$ \cite{AKS01}.
On the other hand, Aharonov and Regev~\cite{AR04} prove than an
$O(\sqrt n)$-approximation  can be
computed in $\np \cap co\np$,\footnote{We have not quite defined
what it means to have an approximation algorithm in $\np \cap co\np$.
See \cite{AR04} for a discussion.} and
so the problem of finding $O(\sqrt n)$-approximate solutions
cannot be \np-hard unless $\np=co\np$.

Interest in the SVP is motivated in part by a result of Ajtai \cite{A96}, showing that if
$n^c$ approximation of SVP cannot be done in polynomial time, for a certain
constant $c$, then there is a problem in \np\ that does not have
any algorithm running in polynomial time on  average. Ajtai's result
was the first one to show that a problem in \np\
is hard on average provided that another problem in \np\ is hard on worst-case.
Even more interestingly, Ajtai and Dwork \cite{AD97} presented a public key
cryptosystem that is secure provided that a version of SVP is hard to approximate. 
The results of Ajtai and Dwork \cite{A96,AD97} have been later improved by 
 Regev \cite{R03} and by Micciancio and Regev~\cite{MR04}. Currently, the cryptosystems
and average-case complexity results of \cite{A96,AD97,R03,MR04}
depend on the worst-case complexity of
problems known to be in $\np\cap co\np$, but it is conceivable that the analyses
could be  improved so that they rely on problems that are \np-hard to
approximate. If so, there would be average-case hard problems, one-way functions and
even private-key encryption be based on the assumption that $\p\neq\np$
(or, more precisely, that $\np\not\subseteq \bpp$). Towards this goal, one needs,
of course, strong inapproximability results for SVP.

The first inapproximability result for SVP was proved by Micciancio \cite{M01},
who showed that there is no $(\sqrt 2-\epsilon)$-approximate algorithm for SVP for
any $\epsilon>0$, unless $\rp = \np$. This  result has been substantially improved
by Khot~\cite{K04}, who proves that 
no $O(1)$-approximate algorithm  for SVP exists unless $\rp=\np$, and that no
$2^{(\log n)^{1/2-\epsilon}}$-approximate algorithm exists unless $\np \subseteq \qrp$,
for every $\epsilon>0$.

\subsubsection*{Closest Vector Problem}

In the {\em Closest Vector Problem}, abbreviated CVP, we are given a lattice
and a vector, and the problem is to find the lattice element that is closest
(in $\ell_2$ norm)
to the given vector. The approximation algorithms that are known for the
SVP can be adapted to approximate CVP within the same ratios~\cite{LLL82,S87,AKS01},
and the techniques of Aharonov and Regev also show that achieving an
 $O(\sqrt n)$ approximation is in $\np \cap co\np$ \cite{AR04}.

The best inapproximability result for CVP is due to Dinur et al., who show that there can be
no $n^{1/O(\log\log n)}$-approximate algorithm unless $\p=\np$
\cite{DKRS03}.

\subsubsection*{Other Norms}

Interesting variants of the SVP and CVP  arise when we measure the 
length of vectors using other $\ell_p$ norms, where the $\ell_p$
norm of a vector $\bv = (v_1,\ldots,v_n)$ 
is $|| \bv||_p = (\sum_i |v_i|^p)^{1/p}$; the problems
are then denoted $\svp_p$ and $\cvp_p$, respectively. Of interest
is also the case of the $\ell_\infty$ norm $|| v||_\infty = \max_i |v_i|$,
that defines the problems $\svp_\infty$ and $\cvp_\infty$.

Any $\ell_p$ norm is a $O(\sqrt n)$ approximation of the $\ell_2$ norm, so the
exponential approximation of \cite{LLL82,S87,AKS01} also apply to other $\ell_p$ norm.

For $\svp_p$,
Micciancio's result \cite{M01} proves a hardness
of approximation within $2^{1/p}-\epsilon$ for any $\ell_p$
norm and $\epsilon>0$.  For every $\delta>0$ and for every
sufficiently large $p$,  Khot \cite{K03}
proves hardness $p^{1-\delta}$. The new result of Khot
\cite{K04} rules out $r$-approximate algorithms for
$\svp_p$ for every constant $r$ and every  $1< p<\infty$,
assuming $\rp\neq\np$, and rules out $2^{(\log n)^{1/2-\epsilon}}$-approximate
algorithm for every $1< p<\infty$ and $\epsilon>0$, assuming $\np\not\subseteq \qrp$.
Micciancio's result remains the strongest known one for the $\ell_1$ norm.

For $\cvp_p$, the hardness result
of Dinur and others \cite{DKRS03}, ruling out $n^{1/O(\log\log n)}$-approximate
algorithms, assuming $\p\neq\np$ applies
to any $\ell_p$ norm $1\leq p < \infty$.

Regarding the $\ell_\infty$ norm, Dinur~\cite{D02}
proves that there is  no $n^{O(1/\log\log n)}$-approximate algorithm for $\svp_\infty$
and $\cvp_\infty$, assuming $\p\neq\np$.
 
\subsection{Decoding Linear Error-Correcting Codes}

Let $\F$ be a field and $A$ be a $n\times k$ matrix over $\F$.
The mapping $C(x) = A  \cdot x$ is a linear error-correcting code
with minimum distance $d$ if for every two disting inputs
$x\neq x' \in \F^k$ the vectors $C(x)$ and $C(x')$ differ
in at least $d$ entries. It is easy to see that this happens
if and only if for every $x\neq \bfz$ the vector $C(x)$
has at least $d$ non-zero entries. In turn, this happens
if and only if every $d$ rows of $A$ are linearly independent.

Error-correcting codes are used for the transmission
(and storage) of information using unreliable media.
Suppose that a sender has a message $x\in \F^k$ and
that he wants to send it to a recipient using an unreliable
channel that introduces transmission errors, and suppose
also that the sender and the recipient know a
linear error-correcting code $C:\F^k\to \F^n$ of minimum distance $d$.
Then the sender can compute $C(x)$ and send it to the recipient
through the channel. Due to transmission errors, the recipient
receives a string $y\in \F^n$, that may be different from $C(x)$.
Assuming that fewer that $d/2$ errors occurred, there is a unique
string in the range of $C()$ that is closest to $y$, and this
string is $C(x)$. If there is confidence that no more than $d/2$
errors occurred, then the recipient can reconstruct $x$ this way.

Clearly, the larger is $d$ the higher is the reliability of
the system, because more transmission errors can be tolerated.
Also, for fixed $d$ and $k$, the smaller is $n$ the more efficient
is the transmission. Therefore, we would like to
 construct codes, that is, matrices, with large
$d$ and small $n$ for given $k$. Asymptotically, we would like to find
large $\rho$, $\delta$ such that for every $n$
$k= \rho n$ and $d = \delta n$. It is beyond the scope of this
paper to give an overview of known results about this problem.
The reader is referred to a coding theory book
like \cite{vanLint99} for more information.

The natural optimization problems associated with linear codes
are the following:
Suppose we are given a linear code, specified
by the encoding matrix $A$, can we compute
its minimum distance? And given a good code $C$ of large
minimum distance and a string $y$, can
we find the  decoding $x$ such that $C(x)$ is closest to $y$?

The first problem is similar to SVP, and it is formally defined
as follows: given a matrix $A\in \F^{n\times k}$, where $\F$
is a field, find the non-zero vector $x\in \F^k$ such
that $A\cdot x$ has the smallest number of non-zero entries.
Dumer, Micciancio and Sudan~\cite{DMS03} prove that,
assuming that $\np\not\subseteq \qrp$,
the problem cannot be approximated within
a factor $2^{(\log n)^{1-\epsilon}}$.
Their result holds even if the fielf $\F$ is restricted to be $\B$.
The second problem, Nearest Codeword,
is similar to CVP. Formally, our
input is  a matrix $A\in \F^{n\times k}$ and a vector $y\in \F^n$,
and the goal is to find a vector $x\in \F^k$ such that
the number of entries where $A\cdot x$ and $y$ differ is minimized.
Arora et al.~\cite{ABSS93} prove that the problem
cannot be approximated within a factor $2^{(\log n)^{1-\epsilon}}$,
assuming $\np\not\subseteq \qp$.

\subsection{The Traveling Salesman Problem}

In the Traveling Salesman Problem (TSP) we are given a complete undirected
graph $G=(V,E)$ with non-negative weights $w:E\to \R$
on the edges. We think of the vertices as being ``cities''
and the weight of the edge $(u,v)$ as the ``distance'' between 
two cities.
The goal is to find a route that touches
each city exactly once and that has minimal total length.

If the distances are allowed to be arbitrary, then it is easy
to see that even deciding whether the optimum is zero or not is
an \np-hard problem, and so no approximation is possible \cite{SG76}.
One, however, is typically interested in instances where the
distances satisfy the triangle inequality, and thus
describe a metric on the set of cities. This special
case is called the Metric TSP, or $\Delta$TSP. The metric TSP is equivalent
to the variation of TSP with arbitrary distances in which
the  route is required to pass through each city {\em at least} once, but
it is allowed pass more than once through some city.

Almost 30 years ago, Christofides devised a
$3/2$-approximate algorithm for Metric TSP~\cite{C76}, and
the performance ratio of his algorithm has never been improved
upon.

Papadimitriou and Yannakakis \cite{PY93:tsp} present an approximation-preserving
reduction from Max 3SAT-$d$ to Metric TSP. The reduction, together with other
reductions seen in the previous section and with the PCP Theorem, implies
that there is no PTAS for Metric TSP unless $\p=\np$. More specifically,
there is a constant $\epsilon_{TSP}>0$ such that if there is a polynomial
time $(1+\epsilon_{TSP})$-approximate algorithm for Metric TSP then $\p=\np$.
The value of $\epsilon_{TSP}$ implied by the PCP Theorem and the reductions
is extremely small, and there has been work devoted to derive stronger
inapproximability results.
 Papadimitriou and Vempala \cite{PV00}, using Theorem \ref{th:haastad}
(H\aa stad's three-query PCP construction \cite{H97}) and a new reduction
show that there is no polynomial time algorithm for Metric TSP
with performance ratio better that $220/219$, unless $\p=\np$.

Grigni, Koutsopias and Papadimitriou show that Metric  TSP has a PTAS
in the special case in which the metric is the shortest path metric of
a planar graph \cite{GKP95}.
Arora \cite{A98} and Mitchell \cite{M99} show that there is a PTAS in the
special case in which cities are points in $\R^2$ and distances are
computed according to the $\ell_2$ norm. Arora's PTAS also works for other
$\ell_p$ metrics and in higher dimension,
with running time doubly exponential in the number
of dimensions. Trevisan \cite{T97} shows that, if $\p\neq\np$, then 
Metric TSP in $\R^{d(n)}$ with an $\ell_p$ metric has no PTAS if 
$d(n) = \Omega(\log n)$. This implies that the double exponential dependency 
of the running time on the number of dimensions in Arora's PTAS
cannot be improved unless every problem in \np\ has subexponential algorithms.

Another interesting version is Asymmetric TSP, in which the ``distance''
from a city $u$ to a city $v$ may be different from the distance
from $v$ to $u$. The Asymmetric $\Delta$TSP is the restriction of Asymmetric TSP
to asymmetric ``distance'' functions that satisfy the  triangle inequality.
A polynomial $O(\log n)$-approximate algorithm for Asymmetric $\Delta$TSP
is known \cite{FGM82}. Papadimitriou and Vempala \cite{PV00} prove
hat there is no polynomial time algorithm for Metric TSP
with performance ratio better that $117/116$,\footnote{The
conference version of \cite{PV00} claimed that no polynomial
$(129/128-\epsilon)$-approximate algorithm for Metric TSP
and no $(42/41-\epsilon)$-approximate algorithm for Asymmetric $\Delta$TSP
 exists for any $\epsilon>0$, assuming $\p\neq\np$. The stronger claim
was due to an error in the proof. The corrected version of the paper,
with the weaker inapproximability results, is available from the authors.}
unless $\p=\np$. It remains an interesting open question to rule
out constant factor approximation for Asymmetric $\Delta$TSP, or
to devise a constant-factor approximation algorithm.

\subsection{Coloring Problems}
\label{sec:color}

In the graph coloring problem we are given an undirected 
graph $G=(V,E)$ and we want to label each vertex with
a ``color'' so that every edge has endpoints of different
colors. The goal is to use as few colors as possible.

Feige and Kilian prove that, if $\zpp\neq\np$, then, for every
$\epsilon$, there is no $n^{1-\epsilon}$-approximate algorithm
for this problem, where $n$ is the number of vertices.

An interesting special case of the problem is to devise algorithms
that color a 3-colorable graph with a minimum number of colors.
(On other graphs, the algorithm may have an arbitrary behavior.)
Blum and Karger~\cite{BK97} (improving previous work by
Karger, Motwani and Sudan~\cite{KMS98} and by Wigderson~\cite{W83}), prove
that there is a polynomial time algorithm that colors every 3-colorable
graph with at most $n^{3/14} \cdot \poly\log(n)$ colors.
Khanna, Linial and Safra \cite{KLS93} prove that there is no polynomial
time algorithm that colors every 3-colorable graph using at most four colors,
unless $\p=\np$, and this is still the strongest known inapproximability
result for this problem. This is one of the largest gaps
between known approximation algorithms and known inapproximability
results for a natural and well-studied problem.

It appears that methods based on probabilistically checkable proofs
could be applied to the problem of coloring 3-colorable graphs, 
provided that one uses  a different defnition of ``soundness.''
Guruswami, H\aa stad and Sudan~\cite{GHS02}
describe a type of
PCP that could be used to prove hardness of 3-coloring.
They are able to construct such PCPs with parameters that are sufficient
to prove intractability for the {\em Hypergraph} Coloring problem.
In the Hypergraph Coloring problem
we are given a hypergraph $H=(V,E)$, where $E$ is a collection
of subsets of $V$, and we want to color the vertices with a minimum number
of colors so that there is no edge that contains vertices all of the same
color. A hypergraph is $r$-uniform if every hyperedge contains exactly
$r$ vertices. An undirected graph is a $2$-uniform hypergraph.
Guruswami et al.~\cite{GHS02}
show that, if $\p\neq\np$, there is no polynomial time algorithm that, given 
a 2-colorable 4-uniform hypergraph, finds a feasible coloring that uses a
constant number of colors.
Dinur, Regev and Smyth~\cite{DRS02} (see also \cite{K02}) prove the same result for 3-uniform
hypergraphs.

Regarding $k$-colorable graphs for large constant $k$, Khot~\cite{K01}
proves that there is no polynomial time algorithm that, given
a $k$-colorable graph, colors it with $k^{O(\log k)}$ colors.

\subsection{Covering Problems}
\label{sec:cover}

Using H\aa stad's PCP construction (Theorem \ref{th:haastad}) and the
reduction of \cite{FGLSS96}, one can prove that there is no polynomial
time algorithm for the Vertex Cover problem with performance ratio better
than $7/6$.

This result has been improved by Dinur and Safra~\cite{DS02}, who prove that
there is no approximation  algorithm for Vertex Cover with
performance ratio smaller than $10\sqrt 5 - 21 = 1.36\ldots$- unless $\p=\np$.
A simple 2-approximate algorithm has been known to exist for thirty years \cite{G74}.
If the unique games conjecture is true, then 2 is the best possible performance
ratio for polynomial time algorithms, 
unless $\p=\np$ \cite{KR03}. (See also Section \ref{sec:uniquegame} below.)

In the Set Cover problem we are given a collection of subsets
$S_1,\ldots,S_m$ of a set $U$ such that $\bigcup_i S_i = U$.
The goal is to find a smallest set $I\subseteq \{1,\ldots,m\}$
such that $\bigcup_{i\in I} S_i = U$. The problem can be
approximated within a factor $\ln |U| +1$,  using either a greedy or a primal-dual 
algorithm \cite{J74,L75,C79}. More precisely, if
every set is of size at most $k$, then the approximation
ratio of the algorithm is at most $H_k$, defined
as $H_k = 1+1/2+\ldots+1/k \leq 1 + \ln k$. Feige~\cite{F98}
shows that there cannot be a $(1-\epsilon)\ln |U|$ approximate
algorithm, for any $\epsilon>0$, unless $\np\subseteq \qp$. 
Trevisan~\cite{T01} notes that Feige's proof also implies that
there is a constant $c$ such that the Set Cover problem with sets of size $k$ 
(where $k$ is constant) has no  $(\ln k -c\ln\ln k)$-approximate algorithm
unless $\p=\np$.

In the Hitting Set problem we are given  collection of subsets
$S_1,\ldots,S_m$ of a set $U$.
The goal is to find a smallest set $S\subseteq U$
such that $S_i \cap S \neq \emptyset$ for every $i=1,\ldots,m$.
It is easy to see that the Set Cover problem and the Hitting
Set problem are equivalent, and that there is an $r(k)$-approximate
algorithm for the Set Cover problem, where $k$ is the size of the largest set,
if and only if there is an $r(t)$-approximate algorithm for the Hitting
Set problem, where $t$ is an upper bound to the number of sets to which an
element of $U$ may belong. Similarly, there is an $r(t)$-approximate
algorithm for the Set Cover problem, where $t$ is an upper bound to the number of sets to which an
element of $U$ may belong if and only if there is an $r(k)$-approximate algorithm for the Hitting
Set problem, where where $k$ is the size of the largest set. 
In particular, it follows from the results of \cite{J74,L75,C79,F98}
that there is a $(1+\ln m)$-approximate algorithm for Hitting Set,
and that there is no $(1-\epsilon)\cdot \ln m$-approximate algorithm
with $\epsilon>0$ unless $\np\subseteq \qp$.

Consider now the Hitting Set problem restricted to instances where
every set is of size at most $k$. This is equivalent to the Set Cover
problem restricted to instances where every element of $U$ may appear
in at most $k$ sets. Another equivalent formulation of the problem
is as 
the Vertex Cover problem in $k$-uniform hypergraphs. (In particular, it
is just  the vertex cover problem in graphs when $k=2$.) A simple
generalization of the 2-approximate algorithm for Vertex Cover
yields a $k$-approximate algorithm for this problem \cite{H82}. Dinur et al.~\cite{DGKR03}
prove  there is no $(k-1-\epsilon)$-approximate algorithm, $\epsilon>0$,
 unless $\p=\np$. Assuming that the unique games conjecture holds,
even $(k-\epsilon)$-approximate algorithms are impossible \cite{KR03}.

\section{Other Topics}
\label{sec:misc}

\subsection{Complexity Classes of Optimization Problems}

Prior to the discovery of the connection between probabilistically
checkable proofs and inapproximability, research on inapproximability
focused on the  study of approximation-preserving reductions among problems,
and on classes of optimization problems and completeness results.

The first definitions of approximation-preserving reductions appear in work by 
Paz and Moran \cite{PM81} and by Ausiello, D'Atri and Protasi \cite{ADP80}.

Orponen and Manilla \cite{OM87} define the class NPO, and show that
certain (artificial) problems are NPO-complete under approximation-preserving
reductions. Kann \cite{K93} finds problems that
are complete for NPO-PB, the restriction of NPO to problems whose objective function
is polynomially bounded in the length of the input.
Crescenzi and Panconesi \cite{CP91} study the class APX of
problems that admit an $O(1)$-approximate algorithm. They show the existence
of (artificial) complete problems in APX with respect to reductions
that preserve the existence of approximation scheme. It is easy to see
that, if $\p\neq\np$, there are problems in APX that do not have approximation
schemes, and so, in particular, no APX-complete problem can have a PTAS
unless $\p=\np$. These results were meant as proofs-of-concept for the
program of proving inapproximability results by proving completeness results
under approximation-preserving reductions. None of these papers, however,
proves inapproximability results for natural problems.

A very different, and more productive, direction was taken by Papadimitriou and Yannakakis \cite{PY91}. They follow the intuition that Max 3SAT should be as central a problem to the study of 
constant
factor approximation as 3SAT is to the study of decision problems in NP. 
Papadimitriou and Yannakakis identify a class of optimization problems, called Max SNP,
that can be defined using a certain fragment of logic. They show that
Max SNP  includes Max 3SAT
and other constraint satisfaction, and they prove that Max 3SAT is complete
for Max SNP  under approximation-preserving reductions. It follows that Max 3SAT has
a PTAS if and only if every problem in Max SNP has a PTAS.
More importantly, Papadimitriou and Yannakakis present several
 reductions from Max 3SAT using approximation-preserving reductions,
 including ones that we presented in earlier
sections. The reductions proved in \cite{PY91}, and in other papers such as
\cite{BP89,PY93:tsp}, implied that if one could rule out approximation schemes for
Max 3SAT (or, equivalently, for any other problem in Max SNP), then inapproximability
results would follow for many other problems. As we discussed earlier, the PCP Theorem
finally proved the non-existence of approximation schemes for Max 3SAT (assuming
$\p\neq\np$), and the reductions of \cite{PY91,BP89,PY93:tsp} played the role that
they were designed for. Around the same time, in another interesting result,
Berman and Schnitger \cite{BS92} proved that if Max 2SAT does not have a probabilistic PTAS (or equivalently, in light
of \cite{PY91}, if Max 3SAT does not have a probabilistic PTAS), 
then there is a $c>0$ such that there is no $n^c$-approximate
algorithm for  Independent Set. The reduction of Berman and Schnither and the PCP Theorem
imply that there is a $c>0$ such that there is no $n^c$-approximate
algorithm for Indepedent Set, assuming $\rp\neq\np$. As we discussed in Section \ref{sec:is},
Arora et al.~\cite{ALMSS92} achieve the same conclusion from the assumption
that $\p\neq\np$, by improving the soundness of the PCP construction using random walks
on an expander. As discussed in \cite{AFWZ95}, one can see the reduction
of Arora et al. \cite{ALMSS92} as a ``derandomized'' version of the reduction
of \cite{BS92}.

The most interesting results in the paper of Papadimitriou and Yannakakis \cite{PY91}
were the approximation preserving reductions from Max 3SAT to other problems, but it
was the idea of defining classes of optimization problems using fragments
of logical theories that caught other people's immagination.
Other papers followed with logic-based definitions of optimization classes
having natural complete problems, such as the work of
Kolaitis and Thakur \cite{KT94,KT95} and of  Panconesi and Ranjan \cite{PR90}.

The connection between proof checking and approximation and the PCP theorem
allowed researchers to prove inapproximability results directly,
without using notions of completeness, and gave central importance
to reductions between specific problems.

Khanna, Motwani, Sudan and Vazirani \cite{KMSV94} revisited the issue of
completeness in approximation classes, 
and gave a definitive treatment. Khanna et al. \cite{KMSV94}
 show how to use PCP to prove natural
problems complete in their respective classes. For example they show that
Max 3SAT is complete in APX, and that Independent Set is complete in
poly-APX, the class of problems that are $n^c$-approximable for some $c>0$,
where $n$ is the length of the input. Crescenzi et al.~\cite{CKST99}
address some finer points about what definition of reduction should be used
to prove general completeness results.

An interesting related line of work, developed by Chang and others~\cite{CGL94,C94,CKST99},
is a characterization of the complexity of approximation problems in terms
of ``query complexity'' to \np\ oracle. To see the point of this work, consider
for example the problem of approximating Max 3SAT within a $1.1$ versus a $1.01$
factor. Both problems are \np-hard, and both problems can be solved in polynomial
time if $\p=\np$, so in particular if we had a polynomial time $1.1$ approximate
algorithm for Max 3SAT we could also design a  polynomial time $1.01$-approximate
algorithm that uses the $1.1$-approximate one as a subroutine. 
If we follow the proof of this result, we see that the hyptothetical 
$1.1$-approximate algorithm is used more than once as a subroutine. 
Is this necessary? The answer
turns out to be YES. Roughly speaking, computing a $1+\epsilon$ approximation is equivalent
to answering  $\Theta(1/\epsilon)$ non-adaptive queries to an \np\ oracle, and the power
of what can be computed with $k$ queries to an $\np$ oracle is strictly less than what can be done
with $k+1$ queries (unless the polynomial hierarchy collapses). 
In contrast, in Independent Set, for every two constants $1<r_1<r_2$, we can design
an $r_1$-approximate algorithm for Independent Set that uses only once a
an $r_2$-approximate algorithm. This property of the Independent Set
problem (that different constant factor approximations have the same complexity)
is called {\em self-improvability}. From the point of view of query complexity,
computing a constant-factor approximation for Independent Set is equivalent to answering 
$\Theta(\log n)$
queries to an \np\ oracle. A consequence of these characterizations of
approximation problems using query complexity is, for example, that one can prove that the
Set Cover problem is not self-improvable unless the polynomial
hierarchy collapses \cite{CKST99}, a result that it is not clear how to prove
without using this machinery.

\subsection{Average-Case Complexity and Approximability}

Feige \cite{F02} proposes to use the conjectured average-case
complexity of certain distributional problems\footnote{A {\em distributional
problem} is  a decision problem equipped with a probability distribution
over the inputs.} in order to prove inapproximability
results. More specifically, Feige consider the 3SAT problem,
together with the distribution where an instance with $n$ variables
and $cn$ clauses, with $c$ being a large constant, is picked uniformly
at random from all instances with $n$ variables and $cn$ clauses.
For sufficiently large constant $c$, such instances have an extremely
high probability of being unsatisfiable, but is seems difficult
to {\em certify} that specific instances are unsatisfiable. 
Feige considers the assumption\footnote{He does not formulate it as a conjecture.}
that, for every choice of $c$, there is no polynomial time algorithm
that on input a satisfiable formula never outputs ``UNSAT,'' and that
on input a formula picked from the above distribution outputs ``UNSAT''
with high probability. (The probability being computed over the
distribution of formulas.)

Under this assumption, Feige shows that there is no $(4/3-\epsilon)$-approximate
algorithm for  the {\em Minimum Bisection} problem, for any $\epsilon>0$. 
The  Minimum Bisection problem is defined as follows: we are given an undirected graph $G=(V,E)$,
with $V$ having even cardinality, and our goal is to find a partition of
$V$ into two subsets $S,\bar S$ of equal size such that a minimum number
of edges crosses the partition. An approximation algorithm
 due to 
Feige and Krautghamer \cite{FK02}, and it achieves a factor
of $O((\log n)^2)$. This has been recently improved to $O((\log n)^{1.5})$
by Arora, Rao and Vazirani \cite{ARV04}.
No inapproximability result was known for this
problem prior to \cite{F02}. Khot \cite{K04:bisection} recently
proved that the Minimum Bisection problem does not have an approximation
scheme assuming that $\np$ is not contained in sub-exponential time.
More precisely, Khot shows that for every $\epsilon>0$ there is
a constant $c_\epsilon >0$ such that there is no $(1+c_\epsilon)$-approximate
algorithm for Minimum Bisection unless 3SAT can be solved in time
$2^{O(n^\epsilon)}$.

Alekhnovich  \cite{A03} defines other distributional problems (related to E3LIN)
and shows that certain problems in coding theory are hard to approximate provided
that the distributional problem is hard on average.

\subsection{Witness Length in PCP Constructions}

The proof of the PCP theorem shows that for every NP-complete problem,
for example 3SAT, we can encode a witness in such a way that it can be
tested with a constant number of queries. How long is that encoding?
The original proof of the PCP theorem shows that
the size of the encoding is 
polynomial in the number of variables and clauses of the formula.
Improvements in~\cite{PS94,FS95} shows that one can achieve cubic
encoding length in the PCP Theorem with a moderate query complexity,
or encoding length $N^{1+\epsilon}$, where $N$ is the size of the formula,
using query complexity that depends on $\epsilon$.
Recent work shows that the witness needs only be of
nearly linear~\cite{HS00,BSVW03,BGHSV04} length in the size of the formula.
It is conceivable that the witness could be of size linear in the
number of clauses of the formula.

The question of witness length in PCP construction is 
studied mostly for its connections to coding theory, but it has an
interesting connection to approximability as well.

If we let $m$ denote the number of clauses and $n$ the number of variables
in a 3SAT instance, it is believed that the satisfiability problem
 cannot be solved in time $2^{o(n)}$, where
$n$ is the number of variables, even in instances where $m=O(n)$.\footnote{In fact,
Impagliazzo, Paturi and Zane \cite{IPZ01} prove that, roughly speaking,
if 3SAT has a $2^{o(n)}$ time algorithm that works only on instances where
$m=O(n)$, then there is also a $2^{o(n)}$ time algorithm for 3SAT that works
on all instances.} Suppose now that there were a proof of the PCP Theorem
with a witness of linear size. Then this can be used to show that there
is some $\epsilon>0$ such that a $(1+\epsilon)$-approximate algorithm
for Max 3SAT running in $2^{o(n)}$ time implies an algorithm that
solves 3SAT in $2^{o(n)}$ time. Since such a consequence is considered 
unlikely, we would have to conclude that no sub-exponential algorithm
can even achieve a good approximation for Max 3SAT. Conversely,
a $(1+o(1))$-approximate algorithm for Max 3SAT running in $2^{o(n)}$ time 
would provide some evidence that the proof length in the PCP Theorem cannot
be linear.

\subsection{Typical and Unusual Approximation Factors}
\label{sec:unusual}

Looking at the inapproximability results proved in the 1990s
using  the PCP Theorem, it is easy to see a certain pattern. 
For several problems, such as Max 3SAT, Vertex Cover, Metric TSP,
Metric Steiner Tree, Max CUT, we know that a polynomial time
constant factor approximation exists, and there is an inapproximability
result proving that for some $\epsilon >0$ there is no $(1+\epsilon)$-approximate
algorithm. For other problems, like Clique, Independent Set, and Chromatic
Number, we can prove that there is an $c>0$ such no $n^c$-approximate
algorithm exists, where $n$ is the size of the input. Then there are problems
like Set Cover that are approximable within an $O(\log n)$ factor but not
within a $o(\log n)$ factor, and finally there are some problems
for which we can show an inapproximability result of the form $2^{(\log n)^c}$
for some constant $c>0$, although we believe that the right lower bound
is $n^c$ for some $c>0$. 

For maximization problems, known results either prove that no PTAS exists (while a constant factor
approximation is known), or that a $n^c$ approximation is impossible,
for some $c>0$. For minimization problems, $\Omega(\log n)$ inapproximability
results are also common, as well as inapproximability results of the form $2^{(\log n)^c}$.

Arora and Lund \cite{AL96} notice this pattern in their survey paper and
they classify optimization problems in four classes (called Class I to Class IV)
according to the known inapproximability result.

Of course, a priori, there is no reason why the approximability of natural
optimization problem should fall into a small number of cases, and, in fact,
unless $\p=\np$, for every (efficiently computable) function $f(n)$
there are (artificial) problems that can be approximated within $f(n)$
but no better. The interesting question is whether the pattern
is an effect of our proof techniques, or whether the approximability of
natural problems tends indeed to fall into one of a few typical cases.
The latter possibilities has interesting precedents in complexity theory:
for example almost all the natural decision problems in \np\ are either
known to be solvable in polynomial or known to be \np-complete, even though,
if $\p\neq\np$, there are (artificial) problems that have an infinite variety
of intermediate complexities \cite{Lad75}. Schaefer~\cite{Sch78} found
in 1978 an interesting ``explanation'' for the latter phenomenon. He considers
an infinite class of boolean satisfiability problems in \np, and he shows that
each such problem is either in \p\ or \np-complete.  
Creignou \cite{C95}, and Khanna, Sudan, Trevisan and Williamson \cite{KSTW00}
(see also the monograph \cite{CKS01})
apply the same framework to optimization problems, and definite
infinite classes of minimization of maximization problems that can be
expressed in terms of boolean constraint satisfaction. They show that
maximization problems in the class are either solvable in polynomial
time, or can be approximated within a constant factor but not
to arbitrary good constant factors (Class I in the classification of \cite{AL96}),
or are hard to approximate within a factor $n^c$ for some $c>0$
(Class IV), or it is intractable to even find a feasible solution.
Minimization problems are either solvable in polynomial time, or
fall into one of five equivalence classes (two corresponding to Class II,
two corresponding to Class III and one corresponding to Class IV), or
it is intractable to find a feasible solution.
It is remarkable that the inifinite class of problems studied in \cite{C95,KSTW00}
would show only a constant number of distinct approximability behaviour,
and that these cases would allign themselves with the classification of \cite{AL96}.

Recent results, however, have shown that some natural problems
have approximation thesholds that do not fit the above scheme.

For example, Halperin   and Krauthgamer \cite{HK03} prove a $\Omega((\log n)^{2-\epsilon})$
inapproximability result for  the Group Steiner Tree problem for every $\epsilon>0$,
a negative result that applies even to the restriction of the problem to trees.
The problem can be approximated within a $O((\log n)^2)$ factors in trees~\cite{GKR00}
and $O((\log n)^3)$ in general graphs~\cite{Bar96,FRT03}. This result shows that
there are natural problems for which the best possible approximation is super-logarithmic
but polylogarithmic.

Chuzhoy and Naor \cite{CN04} prove an $\Omega(\log \log n)$ inapproximability
result for the Min-congestion Unsplittable Flow problem. An
$O(\log n/\log \log n)$-approximate algorithm by Raghavan and Thompson \cite{RT87}
is known for this problem. This result shows that there are natural problems
for which the best possible approximation is sub-logarithmic but super-constant.

The most surprising result along these lines is the recent proof by
Chuzhoy et al.~\cite{CGHKKN04} of a $\log^* k -O(1)$ inapproximability
result for the Asymmetric $k$-Center Problem. An $O(\log^* k)$-approximate
algorithm is known \cite{PV98,Arch01}.

\subsection{Inapproximability Results versus ``Integrality Gaps''}

Several approximation algorithms are based on linear programming
or semidefinite programming relaxations of the problem of interest.
Typically, in order to prove that an algorithm based on a relaxation
is $r$-approximate, one proves the stronger statement that the
algorithm finds a solution whose cost is within a factor $r$ from
the optimum of the relaxation. In particular, this shows that
the optimum of the relaxation is within a factor $r$ from
the optimum of the combinatorial problem.

For example, the Goemans-Williamson approximation algorithm
for Max CUT~\cite{GW94} finds a cut whose cost is at
least $\alpha = .878\ldots$ times the cost of the optimum of the relaxation.
Such an analysis also proves, as a consequence, that the optimum
of the relaxation is always at most  $1/\alpha$ times the
true optimum.

In order to show that such an analysis is tight, one can
try and construct an instance of the combinatorial problem
for which the ratio between the true optimum and the optimum
of the relaxation is as large as the bound implied by
the analysis of the approximation algorithm. For example,
for Max CUT, Feige and Schechtman~\cite{FS02}  prove that
there are graphs for which the ratio between the cost of
 the optimum of the relaxation and the cost  of the maximum
cut is arbitrarily
close to $\alpha$. 

The worst-case ratio between the true optimum of a combinatorial
problem and the optimum of a relaxation is called the {\em integrality
gap} of the relaxation. Research on integrality gaps can be seen
as the study of inapproximability results that apply to a restricted
kind of algorithms (namely, algorithms that use the relaxation
in their analysis). Arora et al.~\cite{ABL02} have recently shown that it is possible to find instances that give large integrality gaps for entire families of relaxations. It would be interesting to generalize their approach to semidefinite programming relaxations.

The study of integrality gap has, moreover, a more direct, if not well understood, relation to inapproximability results.
For example, the inapproximability result for the Set Cover
problem proved by Lund and Yannakakis \cite{LY94}
(and its refinements and improvements in \cite{BGLR93,F98})
relies on a reduction from PCP that involves a family
of sets similar to the one used to prove the integrality
gap of the standard linear programming relaxation of
Set Cover. Similarly, the inapproximability result
for Group Steiner Tree~\cite{HK03} involves a reduction that constructs an instance similar to the one used in~\cite{HKKSW03} to prove an integrality gap. A similar progress from integrality gap to inapproximability result can be see in~\cite{CGHKKN04} for the Asymmetric $K$-Center Problem. Recently, Khot and others \cite{KKMO04} present a PCP construction
that shows that Max CUT cannot be approximate to within a factor small than $\alpha$,
assuming that the unique games conjecture holds  and that the majority function
has certain extremal properties. The PCP construction of \cite{KKMO04} is
stronly inspired by the graphs constructed by Feige and Schechtman~\cite{FS02}
to bound the integrality gap of the Goemans-Williamson relaxation.

Of course it would not be surprising to see that the instances created
by a reduction in an inaproximability result have a large integrality
gap: if not, we would have $\p=\np$. What is surprising is that instances
defined specifically to prove integrality gaps turn out to be useful as
components in reductions proving inapproximability results.

\section{Conclusions}
\label{sec:conclusions}

In this paper we surveyed inapproximability results, a field almost non-existent
in 1990, and then defined by the connection with probabilistic
proof-checking of \cite{FGLSS96,AS92,ALMSS92}. Progress in the years after
the PCP Theorem was
extremely rapid, even outpacing the expectations of the researchers
working in this area. For example, Bellare et al. and Arora \cite{BGS95,A95}
presented two independent arguments that  ``current techniques'' 
could not prove an inapproximability
result stronger than $\sqrt n$ for Independent Set just months before
H\aa stad distributed a preliminary version of his work \cite{H96} 
showing inapproximability within a factor $n^{1-\epsilon}$.
By 1997, optimal inapproximability results were known
for Max 3SAT \cite{H97}, Independent Set \cite{H96}, Coloring \cite{FK96}
and Set Cover \cite{F98}, the problems discussed in
Johnson's paper \cite{J74} that defined the study of approximation algorithms.

After all these successes, the field is still very active and far from mature.
Recent work \cite{DS02,KKMO04} has emphasized the use of higher mathematics
in the analysis of PCP verifier, and of constructions of verifiers that
are so specialised to the application to the problem of interest that
it is hard to draw the line between PCP construction and reduction 
from PCP to problem. The  connection between average-case complexity
and inapproximability is likely to become a major line of work in the future.
It is notable that several important inapproximability
results have been announced just in the last few months, while we were
writing this paper, and are still unpublished at the time
 of writing \cite{K04,K04:bisection,KKMO04}.

Some important open questions seem to still be beyond the reach
of current techniques, and one can look forward to the exciting
new ideas to come in the future to address these questions.
Here we mention some questions that are open at the time
of writing and that the author feels are particularly interesting.
Of course this is only a very small sample, and it is very biased
by the author's interests.

\paragraph{Prove Optimal Results about 2-Query PCPs.}
Several PCP constructions and inapproximability results rely
on versions of the PCP theorem in which the verifier reads only
two entries in the proof, where each entry contains not just
one bit but an element of a larger alphabet of size, say $t=t(n)$.
It is believed\footnote{This conjecture was first made in \cite{BGLR93}.}
 that there is a version of the PCP Theorem with
a 2-query verifier that uses
$O(\log n)$ randomness, accesses a proof written with over an
 alphabet of size $t(n)$, and has soundness $1/t^c(n)$ with $c>0$. 
Raz \cite{R95} proves such a result for constant $t$. For general $t(n)$,
Raz's construction has a verifier that uses $O(t(n)\cdot \log n)$ randomness.
Raz and Safra \cite{RS97} and Arora
and Sudan \cite{AS97} prove a version of the result where the verifier
makes a constant (larger than 2) number of queries, and $t(n)$
is restricted to be $n^{o(1)}$. 

\paragraph{Settle the ``Unique Games Conjecture.''} \label{sec:uniquegame}
The unique games conjecture, formulated 
by Khot~\cite{K02:unique}, is that a certain strong  form of the \pcp\ theorem holds.
The conjecture implies the resolution of various important open questions \cite{K02:unique,KR03},
including the non-existence of $(2-\epsilon)$-approximate algorithms for Vertex Cover
and of $(k-\epsilon)$-approximate algorithm for Vertex Cover in $k$-uniform hypergraphs.
If the unique games conjecture holds, and a certain conjecture about extremal properties 
of monotone functions
is true, then an optimal inapproximability result for Max CUT \cite{KKMO04} would follow,
which would be a remarkable breakthrough.

\paragraph{Prove a Strong Inapproximability Result for Metric TSP.}
The Metric TSP is one of the most well studied optimization problems,
and there is still a very large gap between the known 3/2-approximate algorithm \cite{C76}
and the $220/219$ inapproximability result \cite{PV00}. The reduction of Papadimitriou
and Vempala is an extremely optimized one, and it starts from the optimal
version of the PCP Theorem of H\aa stad \cite{H97}, so it looks like
a very different approach is needed to prove, say, a $1.1$ inapproximability result. 
New results for Vertex Cover \cite{DS02} blur the line
between PCP construction and reduction, and use PCP techniques to directly
construct an instance of the problem of instance. It is possible that
a similar approach could give stronger inapproximability results for Metric TSP.

\paragraph{Make Progress on Graph Partitioning Problems.}
For many problems involving graph partitioning there is a big gap between algorithms
and inapproximability results. Often, no inapproximability results are known at all. 
The {\em sparsest cut}
problem is a typical case of an important graph partitioning problem
for which no inapproximability result is known.
In the {\em sparsest cut} problem we are given an undirected
graph $G=(V,E)$, and the goal is to find
a partition $(S,\bar S)$ of $V$ that
minimizes
\[ \frac {E(S,\bar S)} {\min \{ |S|,|\bar S|\}} \ .\]
\noindent The objective function is defined in
such a way that the optimal solution tends
to be a cut that is both balanced and sparse,
a property that is useful in the design of divide-and-conquer algorithms.

The best known algorithm for Sparsest Cut is 
$O(\sqrt{\log n})$-approximate, where $n$
is the number of vertices, and it is
due to  Arora, Rao and Vazirani \cite{ARV04}, improving
an $O(\log n)$ approximation algorithm by
Leighton and Rao \cite{LR99}. The paper of
Leighton and Rao \cite{LR99} describes various
application of approximate sparsest cut computations
to design divide-and-conquer algorithms.

There is no known inapproximability result for this
problem, and Khot's inapproximability
result for the related problem Min Bisection \cite{K04:bisection}
seem not to generalize to Sparsest Cut. 
Even using average-case assumptions
as in \cite{F02} it is not known how to rule the existence
of a PTAS for Sparsest Cut.

\section*{Acknowledgements}

I thank Vangelis Paschos for giving me the opportunity to write this
survey, and for being very patient with my inability to meet  deadlines.
I am grateful to James Lee and Robi Krautghamer for their very
valuable suggestions that gave shape to Section~\ref{sec:misc}.
I wish to thank Uri Feige, Johan H\aa stad and Subhash
Khot for their comments on an earlier version of this manuscript.

\bibliography{macros,luca}

\newcommand{\etalchar}[1]{$^{#1}$}
\begin{thebibliography}{BSSVW03}

\bibitem[ABL02]{ABL02}
Sanjeev Arora, B\'ela Bollob\'as, and L{\'a}szl{\'o} Lov{\'a}sz.
\newblock Proving integrality gaps without knowing the linear program.
\newblock In {\em Proceedings of the 43rd IEEE Symposium on Foundations of
  Computer Science}, pages 313--322, 2002.

\bibitem[ABSS97]{ABSS93}
Sanjeev Arora, L\'aszl\'o Babai, Jacques Stern, and Z.~Sweedyk.
\newblock The hardness of approximate optima in lattices, codes, and systems of
  linear equations.
\newblock {\em Journal of Computer and System Sciences}, 54(2):317--331, 1997.

\bibitem[ACG{\etalchar{+}}99]{ACGKMP99}
Giorgio Ausiello, Pierluigi Crescenzi, Giorgio Gambosi, Viggo Kann, Alberto
  Marchetti-Spaccamela, and Marco Protasi.
\newblock {\em Complexity and Approximation: Combinatorial Optimization
  Problems and Their Approximability Properties}.
\newblock Springer-Verlag, 1999.

\bibitem[AD97]{AD97}
Mikl{\'o}s Ajtai and Cynthia Dwork.
\newblock A public-key cryptosystem with worst-case/average-case equivalence.
\newblock In {\em Proceedings of the 29th ACM Symposium on Theory of
  Computing}, pages 284--293, 1997.

\bibitem[ADP80]{ADP80}
G.~Ausiello, A.~D'Atri, and M.~Protasi.
\newblock Structure preserving reductions among convex optimization problems.
\newblock {\em Journal of Computer and System Sciences}, 21:136--153, 1980.

\bibitem[AFWZ95]{AFWZ95}
N.~Alon, U.~Feige, A.~Wigderson, and D.~Zuckerman.
\newblock Derandomized graph products.
\newblock {\em Computational Complexity}, 5(1):60--75, 1995.

\bibitem[Ajt96]{A96}
Mikl{\'o}s Ajtai.
\newblock Generating hard instances of lattice problems.
\newblock In {\em Proceedings of the 28th ACM Symposium on Theory of
  Computing}, pages 99--108, 1996.

\bibitem[AK98]{AK98}
N.~Alon and N.~Kahale.
\newblock Approximating the independence number via the $\theta$ function.
\newblock {\em Mathematical Programming}, 80:253--264, 1998.

\bibitem[AKS01]{AKS01}
Miklos Ajtai, Ravi Kumar, and D.~Sivakumar.
\newblock A sieve algorithm for the shortest lattice vector problem.
\newblock In {\em Proceedings of the 33rd ACM Symposium on Theory of
  Computing}, pages 601--610, 2001.

\bibitem[AL96]{AL96}
S.~Arora and C.~Lund.
\newblock Hardness of approximations.
\newblock In {\em Approximation Algorithms for {NP}-hard Problems}. PWS
  Publishing, 1996.

\bibitem[Ale03]{A03}
Michael Alekhnovich.
\newblock More on average case vs approximation complexity.
\newblock In {\em Proceedings of the 44th IEEE Symposium on Foundations of
  Computer Science}, pages 298--307, 2003.

\bibitem[ALM{\etalchar{+}}98]{ALMSS92}
S.~Arora, C.~Lund, R.~Motwani, M.~Sudan, and M.~Szegedy.
\newblock Proof verification and hardness of approximation problems.
\newblock {\em Journal of the ACM}, 45(3):501--555, 1998.
\newblock Preliminary version in {\em Proc. of FOCS'92}.

\bibitem[AR04]{AR04}
Dorit Aharonov and Oded Regev.
\newblock Lattice problems in {NP$\cap$coNP}.
\newblock In {\em Proceedings of the 45th IEEE Symposium on Foundations of
  Computer Science}, 2004.

\bibitem[Arc01]{Arch01}
Aaron Archer.
\newblock Two {$O(\log^* k)$}-approximation algorithms for the asymmetric
  k-center problem.
\newblock In {\em Proceedings of the 8th Conference on Integer Programming and
  Combinatorial Optimization}, pages 1--14, 2001.

\bibitem[Aro94]{A94}
Sanjeev Arora.
\newblock {\em Probabilistic checking of proofs and the hardness of
  approximation problems}.
\newblock PhD thesis, U.C. Berkeley, 1994.

\bibitem[Aro95]{A95}
S.~Arora.
\newblock Reductions, codes, {PCP}'s and inapproximability.
\newblock In {\em Proceedings of the 36th IEEE Symposium on Foundations of
  Computer Science}, pages 404--413, 1995.

\bibitem[Aro98a]{A98:survey}
S.~Arora.
\newblock The approximability of {NP}-hard problems.
\newblock In {\em Proceedings of the 30th ACM Symposium on Theory of
  Computing}, pages 337--348, 1998.

\bibitem[Aro98b]{A98}
Sanjeev Arora.
\newblock Polynomial time approximation schemes for {E}uclidean {Traveling}
  {Salesman} and other geometric problems.
\newblock {\em Journal of the ACM}, 45(5), 1998.

\bibitem[ARV04]{ARV04}
Sanjeev Arora, Satish Rao, and Umesh Vazirani.
\newblock Expander flows and a $\sqrt{\log n}$-approximation to sparsest cut.
\newblock In {\em Proceedings of the 36th ACM Symposium on Theory of
  Computing}, 2004.

\bibitem[AS97]{AS97}
S.~Arora and M.~Sudan.
\newblock Improved low degree testing and its applications.
\newblock In {\em Proceedings of the 29th ACM Symposium on Theory of
  Computing}, pages 485--495, 1997.

\bibitem[AS98]{AS92}
S.~Arora and S.~Safra.
\newblock Probabilistic checking of proofs: A new characterization of~{NP}.
\newblock {\em Journal of the ACM}, 45(1):70--122, 1998.
\newblock Preliminary version in {\em Proc. of FOCS'92}.

\bibitem[Bar96]{Bar96}
Yair Bartal.
\newblock Probabilistic approximations of metric spaces and its algorithmic
  applications.
\newblock In {\em Proceedings of the 37th IEEE Symposium on Foundations of
  Computer Science}, pages 184--193, 1996.

\bibitem[Bel96]{B96}
M.~Bellare.
\newblock Proof checking and approximation: Towards tight results.
\newblock {\em Sigact News}, 27(1), 1996.

\bibitem[BGLR93]{BGLR93}
M~Bellare, S.~Goldwasser, C.~Lund, and A.~Russell.
\newblock Efficient probabilistically checkable proofs and applications to
  approximation.
\newblock In {\em Proceedings of the 25th ACM Symposium on Theory of
  Computing}, pages 294--304, 1993.
\newblock See also the errata sheet in {\em Proc of STOC'94}.

\bibitem[BGS98]{BGS95}
M.~Bellare, O.~Goldreich, and M.~Sudan.
\newblock Free bits, {PCP}'s and non-approximability -- towards tight results.
\newblock {\em SIAM Journal on Computing}, 27(3):804--915, 1998.
\newblock Preliminary version in {\em Proc. of FOCS'95}.

\bibitem[BK97]{BK97}
A.~Blum and D.~Karger.
\newblock An {$\tilde O(n^{3/14})$} coloring algorithm for 3-colorable graphs.
\newblock {\em Information Processing Letters}, 61(1):49--53, 1997.

\bibitem[BP89]{BP89}
M.~Bern and P.~Plassmann.
\newblock The {Steiner} tree problem with edge lengths 1 and 2.
\newblock {\em Information Processing Letters}, 32:171--176, 1989.

\bibitem[BS92]{BS92}
P.~Berman and G.~Schnitger.
\newblock On the complexity of approximating the independent set problem.
\newblock {\em Information and Computation}, 96:77--94, 1992.
\newblock Preliminary version in {\em Proc. of STACS'89}.

\bibitem[BS94]{BS94}
M.~Bellare and M.~Sudan.
\newblock Improved non-approximability results.
\newblock In {\em Proceedings of the 26th ACM Symposium on Theory of
  Computing}, pages 184--193, 1994.

\bibitem[BSGH{\etalchar{+}}04]{BGHSV04}
Eli Ben-Sasson, Oded Goldreich, Prahladh Harsha, Madhu Sudan, and Salil Vadhan.
\newblock Robust {PCP}s of proximity, shorter {PCP}s and applications to
  coding.
\newblock In {\em Proceedings of the 36th ACM Symposium on Theory of
  Computing}, 2004.

\bibitem[BSSVW03]{BSVW03}
Eli Ben-Sasson, Madhu Sudan, Salil~P. Vadhan, and Avi Wigderson.
\newblock Randomness-efficient low degree tests and short {PCP}s via
  $\epsilon$-biased sets.
\newblock In {\em Proceedings of the 35th ACM Symposium on Theory of
  Computing}, pages 612--621, 2003.

\bibitem[CGH{\etalchar{+}}04]{CGHKKN04}
Julia Chuzhoy, Sudipto Guha, Eran Halperin, Sanjeev Khanna, Guy Kortsarz, and
  Joseph Naor.
\newblock Asymmetric k-center is $log^* n$ hard to approximate.
\newblock In {\em Proceedings of the 36th ACM Symposium on Theory of
  Computing}, 2004.

\bibitem[CGL97]{CGL94}
R.~Chang, W.~I. Gasarch, and C.~Lund.
\newblock On bounded queries and approximation.
\newblock {\em SIAM Journal on Computing}, 26(1):188--209, 1997.
\newblock Preliminary version in {\em Proc. of FOCS'93}.

\bibitem[Cha96]{C94}
R.~Chang.
\newblock On the query complexity of clique size and maximum satisfiability.
\newblock {\em Journal of Computer and System Sciences}, 53(2):298--313, 1996.
\newblock Preliminary version in {\em Proc. of Structures'94}.

\bibitem[Chr76]{C76}
N.~Christofides.
\newblock Worst-case analysis of a new heuristic for the travelling salesman
  problem.
\newblock Technical report, Carnegie-Mellon University, 1976.

\bibitem[Chv79]{C79}
V.~Chvatal.
\newblock A greedy heuristic for the set-covering problem.
\newblock {\em Mathematics of Operations Research}, 4:233--235, 1979.

\bibitem[CKS01]{CKS01}
Nadia Creignou, Sanjeev Khanna, , and Madhu Sudan.
\newblock {\em Complexity Classifications of Boolean Constraint Satisfaction
  Problems}.
\newblock SIAM Monographs on Discrete Mathematics and Applications 7, 2001.

\bibitem[CKST99]{CKST99}
P.~Crescenzi, V.~Kann, R.~Silvestri, and L.~Trevisan.
\newblock Structure in approximation classes.
\newblock {\em SIAM Journal on Computing}, 28(5):1759--1782, 1999.

\bibitem[CN04]{CN04}
Julia Chuzhoy and Seffi Naor.
\newblock New hardness results for congestion minimization and machine
  scheduling.
\newblock In {\em Proceedings of the 36th ACM Symposium on Theory of
  Computing}, 2004.

\bibitem[Con93]{C91}
A.~Condon.
\newblock The complexity of the max-word problem and the power of one-way
  interactive proof systems.
\newblock {\em Computational Complexity}, 3:292--305, 1993.
\newblock Preliminary version in {\em Proc. of STACS91}.

\bibitem[Coo71]{C71}
S.A. Cook.
\newblock The complexity of theorem proving procedures.
\newblock In {\em Proceedings of the 3rd ACM Symposium on Theory of Computing},
  pages 151--158, 1971.

\bibitem[CP91]{CP91}
P.~Crescenzi and A.~Panconesi.
\newblock Completeness in approximation classes.
\newblock {\em Information and Computation}, 93:241--262, 1991.
\newblock Preliminary version in {\em Proc. of FCT'89}.

\bibitem[Cre95]{C95}
Nadia Creignou.
\newblock A dichotomy theorem for maximum generalized satisfiability problems.
\newblock {\em Journal of Computer and System Sciences}, 51(3):511--522, 1995.

\bibitem[DGKR03]{DGKR03}
Irit Dinur, Venkatesan Guruswami, Subhash Khot, and Oded Regev.
\newblock A new multilayered pcp and the hardness of hypergraph vertex cover.
\newblock In {\em Proceedings of the 35th ACM Symposium on Theory of
  Computing}, pages 595--601, 2003.

\bibitem[Din02]{D02}
Irit Dinur.
\newblock Approximating {SVP$_\infty$} to within almost-polynomial factors is
  {NP}-hard.
\newblock {\em Theoretical Computer Science}, 285(1):55--71, 2002.

\bibitem[DKRS03]{DKRS03}
Irit Dinur, Guy Kindler, Ran Raz, and Shmuel Safra.
\newblock An improved lower bound for approximating {CVP}.
\newblock {\em Combinatorica}, 23(2):205--243, 2003.

\bibitem[DMS03]{DMS03}
Ilya Dumer, Daniele Micciancio, , and Madhu Sudan.
\newblock Hardness of approximating the minimum distance of a linear code.
\newblock {\em {IEEE} Transactions on Information Theory}, 49(1):22--37, 2003.

\bibitem[DRS02]{DRS02}
Irit Dinur, Oded Regev, and Clifford~D. Smyth.
\newblock The hardness of {3-Uniform} hypergraph coloring.
\newblock In {\em Proceedings of the 43rd IEEE Symposium on Foundations of
  Computer Science}, pages 33--42, 2002.

\bibitem[DS02]{DS02}
Irit Dinur and Shmuel Safra.
\newblock The importance of being biased.
\newblock In {\em Proceedings of the 34th ACM Symposium on Theory of
  Computing}, pages 33--42, 2002.

\bibitem[Fei98]{F98}
Uriel Feige.
\newblock A threshold of $\ln n$ for approximating set cover.
\newblock {\em Journal of the ACM}, 45(4):634--652, 1998.

\bibitem[Fei02a]{F02:icm}
Uriel Feige.
\newblock Approximation thresholds for combinatorial optimization problems.
\newblock In {\em Proceedings of the International Congress of Mathematicians},
  pages 649--658, 2002.
\newblock Volume 3.

\bibitem[Fei02b]{F02}
Uriel Feige.
\newblock Relations between average case complexity and approximation
  complexity.
\newblock In {\em Proceedings of the 34th ACM Symposium on Theory of
  Computing}, pages 534--543, 2002.

\bibitem[FGL{\etalchar{+}}96]{FGLSS96}
U.~Feige, S.~Goldwasser, L.~Lov\'asz, S.~Safra, and M.~Szegedy.
\newblock Interactive proofs and the hardness of approximating cliques.
\newblock {\em Journal of the ACM}, 43(2):268--292, 1996.
\newblock Preliminary version in {\em Proc. of FOCS91}.

\bibitem[FGM82]{FGM82}
A.~Frieze, G.~Galbiati, and F~Maffioli.
\newblock On the worst-case performance of some algorithms for the asymmetric
  traveling salesman problem.
\newblock {\em Networks}, 12:23--39, 1982.

\bibitem[FK94]{FK94}
U.~Feige and J.~Kilian.
\newblock Two prover protocols - low error at affordable rates.
\newblock In {\em Proceedings of the 26th ACM Symposium on Theory of
  Computing}, pages 172--183, 1994.

\bibitem[FK98]{FK96}
Uriel Feige and Joe Kilian.
\newblock Zero knowledge and the chromatic number.
\newblock {\em Journal of Computer and System Sciences}, 57(2):187--199, 1998.

\bibitem[FK02]{FK02}
Uriel Feige and Robert Krauthgamer.
\newblock A polylogarithmic approximation of the minimum bisection.
\newblock {\em SIAM Journal on Computing}, 31(4):1090--1118, 2002.

\bibitem[FRT03]{FRT03}
Jittat Fakcharoenphol, Satish Rao, and Kunal Talwar.
\newblock A tight bound on approximating arbitrary metrics by tree metrics.
\newblock In {\em Proceedings of the 35th ACM Symposium on Theory of
  Computing}, pages 448--455, 2003.

\bibitem[FS95]{FS95}
Katalin Friedl and Madhu Sudan.
\newblock Some improvements to total degree tests.
\newblock In {\em Proceedings of the 3rd Israel Symposium on the Theory of
  Computing and Systems}, pages 190--198, 1995.

\bibitem[FS02]{FS02}
Uriel Feige and Gideon Schechtman.
\newblock On the optimality of the random hyperplane rounding technique for
  {MAX CUT}.
\newblock {\em Random Structures and Algorithms}, 20(3):403--440, 2002.

\bibitem[Gav74]{G74}
F.~Gavril.
\newblock Manuscript cited in \cite{GJ79}, 1974.

\bibitem[GG81]{GG81}
O.~Gabber and Z.~Galil.
\newblock Explicit construction of linear sized superconcentrators.
\newblock {\em Journal of Computer and System Sciences}, 22:407--425, 1981.

\bibitem[GHS02]{GHS02}
Venkatesan Guruswami, Johan H{\aa}stad, and Madhu Sudan.
\newblock Hardness of approximate hypergraph coloring.
\newblock {\em SIAM Journal on Computing}, 31(6):1663--1686, 2002.

\bibitem[GJ76]{GJ76}
M.R. Garey and D.S. Johnson.
\newblock The complexity of near-optimal graph coloring.
\newblock {\em Journal of the ACM}, 23:43--49, 1976.

\bibitem[GJ79]{GJ79}
M.R. Garey and D.S. Johnson.
\newblock {\em Computers and Intractability: a Guide to the Theory of
  {NP}-Completeness}.
\newblock Freeman, 1979.

\bibitem[GKP95]{GKP95}
M.~Grigni, E.~Koutsoupias, and C.H. Papadimitriou.
\newblock An approximation scheme for planar graph {TSP}.
\newblock In {\em Proceedings of the 36th IEEE Symposium on Foundations of
  Computer Science}, pages 640--645, 1995.

\bibitem[GKR00]{GKR00}
Naveen Garg, Goran Konjevod, and R.~Ravi.
\newblock A polylogarithmic approximation algorithm for the group steiner tree
  problem.
\newblock {\em J. of Algorithms}, 37(1):66--84, 2000.

\bibitem[Gra66]{G66}
R.L. Graham.
\newblock Bounds for certain multiprocessing anomalies.
\newblock {\em Bell System Technology Journal}, 45:1563--1581, 1966.

\bibitem[GW95]{GW94}
M.X. Goemans and D.P. Williamson.
\newblock Improved approximation algorithms for maximum cut and satisfiability
  problems using semidefinite programming.
\newblock {\em Journal of the ACM}, 42(6):1115--1145, 1995.
\newblock Preliminary version in {\em Proc. of STOC'94}.

\bibitem[Hal98]{Hal98}
M.~Halldorsson.
\newblock A survey on independent set approximations.
\newblock In {\em APPROX'98}, pages 1--14, 1998.
\newblock LNCS 1444, Springer-Verlag.

\bibitem[H{\aa}s99]{H96}
J.~H{\aa}stad.
\newblock Clique is hard to approximate within $n^{1-\epsilon}$.
\newblock {\em Acta Mathematica}, 182:105--142, 1999.

\bibitem[H{\aa}s01]{H97}
Johan H{\aa}stad.
\newblock Some optimal inapproximability results.
\newblock {\em Journal of the ACM}, 48(4):798--859, 2001.

\bibitem[HK03]{HK03}
Eran Halperin and Robert Krauthgamer.
\newblock Polylogarithmic inapproximability.
\newblock In {\em Proceedings of the 35th ACM Symposium on Theory of
  Computing}, pages 585--594, 2003.

\bibitem[HKK{\etalchar{+}}03]{HKKSW03}
Eran Halperin, Guy Kortsarz, Robert Krauthgamer, Aravind Srinivasan, and Nan
  Wang.
\newblock Integrality ratio for group {Steiner} trees and directed {Steiner}
  trees.
\newblock In {\em Proceedings of the 14th ACM-SIAM Symposium on Discrete
  Algorithms}, pages 275--284, 2003.

\bibitem[Hoc82]{H82}
D.~Hochbaum.
\newblock Approximation algorithms for set covering and vertex cover problems.
\newblock {\em SIAM Journal on Computing}, 11:555--556, 1982.

\bibitem[Hoc96]{Hoc96}
Dorit Hochbaum, editor.
\newblock {\em Approximation Algorithms for {NP}-Hard Problems}.
\newblock PWS Publishing, 1996.

\bibitem[HS85]{HS85}
D.S. Hochbaum and D.B. Shmoys.
\newblock A best possible heuristic for the $k$-center problem.
\newblock {\em Mathematics of Operations Research}, 10(2):180--184, 1985.

\bibitem[HS00]{HS00}
Prahladh Harsha and Madhu Sudan.
\newblock Small {PCP}s with low query complexity.
\newblock {\em Computational Complexity}, 9(3--4):157--201, 2000.

\bibitem[HW01]{HW01}
Johan H{\aa}stad and Avi Wigderson.
\newblock Simple analysis of graph tests for linearity and {PCP}.
\newblock In {\em Proceedings of the 16th IEEE Conference on Computational
  Complexity}, pages 244--254, 2001.

\bibitem[IPZ01]{IPZ01}
Russell Impagliazzo, Ramamohan Paturi, and Francis Zane.
\newblock Which problems have strongly exponential complexity?
\newblock {\em Journal of Computer and System Sciences}, 63(4):512--530, 2001.

\bibitem[IZ89]{IZ89}
R.~Impagliazzo and D.~Zuckerman.
\newblock How to recycle random bits.
\newblock In {\em Proceedings of the 30th IEEE Symposium on Foundations of
  Computer Science}, pages 248--253, 1989.

\bibitem[Joh74]{J74}
D.S. Johnson.
\newblock Approximation algorithms for combinatorial problems.
\newblock {\em Journal of Computer and System Sciences}, 9:256--278, 1974.

\bibitem[Joh92]{J92:pcp}
D.S. Johnson.
\newblock The {NP}-completeness column: An ongoing guide: The tale of the
  second prover.
\newblock {\em Journal of Algorithms}, 13:502--524, 1992.

\bibitem[Kan93]{K93}
V.~Kann.
\newblock Polynomially bounded minimization problems which are hard to
  approximate.
\newblock In {\em Proceedings of 20th International Colloquium on Automata,
  Languages and Programming}, pages 52--63, 1993.

\bibitem[Kar72]{K72}
R.M. Karp.
\newblock Reducibility among combinatorial problems.
\newblock In R.E. Miller and J.W. Thatcher, editors, {\em Complexity of
  Computer Computations}, pages 85--103. Plenum Press, 1972.

\bibitem[Kho01]{K01}
Subhash Khot.
\newblock Improved inaproximability results for maxclique, chromatic number and
  approximate graph coloring.
\newblock In {\em Proceedings of the 42nd IEEE Symposium on Foundations of
  Computer Science}, pages 600--609, 2001.

\bibitem[Kho02a]{K02}
Subhash Khot.
\newblock Hardness results for coloring 3-colorable 3-uniform hypergraphs.
\newblock In {\em Proceedings of the 43rd IEEE Symposium on Foundations of
  Computer Science}, pages 23--32, 2002.

\bibitem[Kho02b]{K02:unique}
Subhash Khot.
\newblock On the power of unique 2-prover 1-round games.
\newblock In {\em Proceedings of the 34th ACM Symposium on Theory of
  Computing}, pages 767--775, 2002.

\bibitem[Kho03]{K03}
Subhash Khot.
\newblock Hardness of approximating the shortest vector problem in high {$L_p$}
  norms.
\newblock In {\em Proceedings of the 44th IEEE Symposium on Foundations of
  Computer Science}, pages 290--297, 2003.

\bibitem[Kho04a]{K04}
Subhash Khot.
\newblock Hardness of approximating the shortest vector problem in lattices.
\newblock In {\em Proceedings of the 45th IEEE Symposium on Foundations of
  Computer Science}, 2004.

\bibitem[Kho04b]{K04:bisection}
Subhash Khot.
\newblock Ruling out {PTAS} for graph min-bisection, densest subgraph and
  bipartite clique.
\newblock In {\em Proceedings of the 45th IEEE Symposium on Foundations of
  Computer Science}, 2004.

\bibitem[KKMO04]{KKMO04}
Subhash Khot, Guy Kindler, Elchanan Mossel, and Ryan O'Donnell.
\newblock Optimal inapproximability results for {MAX-CUT} and other
  two-variable {CSP}s?
\newblock In {\em Proceedings of the 45th IEEE Symposium on Foundations of
  Computer Science}, 2004.

\bibitem[KLS93]{KLS93}
S.~Khanna, N.~Linial, and S.~Safra.
\newblock On the hardness of approximating the chromatic number.
\newblock In {\em Proceedings of the 2nd IEEE Israel Symposium on Theory of
  Computing and Systems}, pages 250--260, 1993.

\bibitem[KMS98]{KMS98}
D.~Karger, R.~Motwani, and M.~Sudan.
\newblock Approximate graph coloring by semi-definite programming.
\newblock {\em Journal of the ACM}, 45(2):246--265, 1998.

\bibitem[KMSV99]{KMSV94}
S.~Khanna, R.~Motwani, M.~Sudan, and U.~Vazirani.
\newblock On syntactic versus computational views of approximability.
\newblock {\em SIAM Journal on Computing}, 28(1):164--191, 1999.
\newblock Preliminary version in {\em Proc. of FOCS'94}.

\bibitem[KR03]{KR03}
Subhash Khot and Oded Regev.
\newblock Vertex cover might be hard to approximate to within $2-\epsilon$.
\newblock In {\em Proceedings of the 18th IEEE Conference on Computational
  Complexity}, 2003.

\bibitem[KSTW00]{KSTW00}
Sanjeev Khanna, Madhu Sudan, Luca Trevisan, and David~P. Williamson.
\newblock The approximability of constraint satisfaction problems.
\newblock {\em SIAM Journal on Computing}, 30(6):1863--1920, 2000.

\bibitem[KT94]{KT94}
P.G. Kolaitis and M.N. Thakur.
\newblock Logical definability of {NP} optimization problems.
\newblock {\em Information and Computation}, 115(2):321--353, 1994.

\bibitem[KT95]{KT95}
P.G. Kolaitis and M.N. Thakur.
\newblock Approximation properties of {NP} minimization classes.
\newblock {\em Journal of Computer and System Sciences}, 50:391--411, 1995.
\newblock Preliminary version in {\em Proc. of Structures91}.

\bibitem[Lad75]{Lad75}
R.~Ladner.
\newblock On the structure of polynomial time reducibility.
\newblock {\em Journal of the ACM}, 22(1):155--171, 1975.

\bibitem[Lev73]{L73}
L.~A. Levin.
\newblock Universal search problems.
\newblock {\em Problemi Peredachi Informatsii}, 9:265--266, 1973.

\bibitem[LLL82]{LLL82}
A.K. Lenstra, H.W. Lenstra, and L.~Lovasz.
\newblock Factoring polynomials with rational coefficients.
\newblock {\em Mathematische Annalen}, 261:515--534, 1982.

\bibitem[Lov75]{L75}
L.~Lovasz.
\newblock On the ratio of optimal integral and fractional covers.
\newblock {\em Discrete Mathematics}, 13:383--390, 1975.

\bibitem[LPS88]{LPS88}
A.~Lubotzky, R.~Phillips, and P.~Sarnak.
\newblock Ramanujan graphs.
\newblock {\em Combinatorica}, 8:261--277, 1988.

\bibitem[LR99]{LR99}
Frank~T. Leighton and Satish Rao.
\newblock Multicommodity max-flow min-cut theorems and their use in designing
  approximation algorithms.
\newblock {\em Journal of the ACM}, 46:787--832, 1999.

\bibitem[LY94]{LY94}
C.~Lund and M.~Yannakakis.
\newblock On the hardness of approximating minimization problems.
\newblock {\em Journal of the ACM}, 41:960--981, 1994.
\newblock Preliminary version in {\em Proc. of STOC'93}.

\bibitem[Mic01]{M01}
Daniele Micciancio.
\newblock The shortest vector problem is {NP}-hard to approximate to within
  some constant.
\newblock {\em SIAM Journal on Computing}, 30(6):2008--2035, 2001.

\bibitem[Mit99]{M99}
J.S.B. Mitchell.
\newblock Guillotine subdivisions approximate polygonal subdivisions: A simple
  polynomial-time approximation scheme for geometric {TSP}, {K-MST}, and
  related problems.
\newblock {\em SIAM Journal on Computing}, 28(4):1298--1309, 1999.

\bibitem[MR04]{MR04}
Daniele Micciancio and Oded Regev.
\newblock Stronger average case to worst case connections.
\newblock Manuscript, 2004.

\bibitem[OM87]{OM87}
P.~Orponen and H.~Mannila.
\newblock On approximation preserving reductions: complete problems and robust
  measures.
\newblock Technical Report C-1987-28, Department of Computer Science,
  University of Helsinki, 1987.

\bibitem[Pap94]{P94}
C.H. Papadimitriou.
\newblock {\em Computational Complexity}.
\newblock Addison-Wesley, 1994.

\bibitem[PM81]{PM81}
A.~Paz and S.~Moran.
\newblock Non deterministic polynomial optimization problems and their
  approximation.
\newblock {\em Theoretical Computer Science}, 15:251--277, 1981.

\bibitem[PR90]{PR90}
A.~Panconesi and D.~Ranjan.
\newblock Quantifiers and approximation.
\newblock In {\em Proceedings of the 22nd ACM Symposium on Theory of
  Computing}, pages 446--456, 1990.

\bibitem[PS94]{PS94}
A.~Polishchuk and D.A. Spielman.
\newblock Nearly-linear size holographic proofs.
\newblock In {\em Proceedings of the 26th ACM Symposium on Theory of
  Computing}, pages 194--203, 1994.

\bibitem[PV98]{PV98}
Rina Panigrahy and Sundar Vishwanathan.
\newblock An {$O(\log^* n)$} approximation algorithm for the asymmetric
  p-center problem.
\newblock {\em J. of Algorithms}, 27(2):259--268, 1998.

\bibitem[PV00]{PV00}
Christos~H. Papadimitriou and Santosh Vempala.
\newblock On the approximability of the traveling salesman problem.
\newblock In {\em Proceedings of the 32nd ACM Symposium on Theory of
  Computing}, pages 126--133, 2000.

\bibitem[PY91]{PY91}
C.~H. Papadimitriou and M.~Yannakakis.
\newblock Optimization, approximation, and complexity classes.
\newblock {\em Journal of Computer and System Sciences}, 43:425--440, 1991.
\newblock Preliminary version in {\em Proc. of STOC'88}.

\bibitem[PY93]{PY93:tsp}
C.H. Papadimitriou and M.~Yannakakis.
\newblock The travelling salesman problem with distances one and two.
\newblock {\em Mathematics of Operations Research}, 18:1--11, 1993.

\bibitem[Raz98]{R95}
R.~Raz.
\newblock A parallel repetition theorem.
\newblock {\em SIAM Journal on Computing}, 27(3):763--803, 1998.
\newblock Preliminary version in {\em Proc. of STOC'95}.

\bibitem[Reg03]{R03}
Oded Regev.
\newblock New lattice based cryptographic constructions.
\newblock In {\em Proceedings of the 35th ACM Symposium on Theory of
  Computing}, pages 407--416, 2003.

\bibitem[RS97]{RS97}
R.~Raz and S.~Safra.
\newblock A sub-constant error-probability low-degree test, and a sub-constant
  error-probability {PCP} characterization of {NP}.
\newblock In {\em Proceedings of the 29th ACM Symposium on Theory of
  Computing}, pages 475--484, 1997.

\bibitem[RT87]{RT87}
P.~Raghavan and C.D. Thompson.
\newblock Randomized rounding: a technique for provably good algorithms and
  algorithmic proofs.
\newblock {\em Combinatorica}, 7:365--374, 1987.

\bibitem[Sch78]{Sch78}
T.J. Schaefer.
\newblock The complexity of satisfiability problems.
\newblock In {\em Proceedings of the 10th ACM Symposium on Theory of
  Computing}, pages 216--226, 1978.

\bibitem[Sch87]{S87}
C.P. Schnorr.
\newblock A hierarchy of polynomial time lattice basis reduction algorithms.
\newblock {\em Theoretical Computer Science}, 53:201--224, 1987.

\bibitem[SG76]{SG76}
S.~Sahni and T.~Gonzalez.
\newblock {P}-complete approximation problems.
\newblock {\em Journal of the ACM}, 23:555--565, 1976.

\bibitem[ST00]{ST00}
A.~Samorodnitsky and L.~Trevisan.
\newblock A {PCP} characterization of {NP} with optimal amortized query
  complexity.
\newblock In {\em Proceedings of the 32nd ACM Symposium on Theory of
  Computing}, 2000.

\bibitem[Sud92]{S92}
M.~Sudan.
\newblock {\em Efficient Checking of Polynomials and Proofs and the Hardness of
  Approximation Problems}.
\newblock PhD thesis, University of California at Berkeley, 1992.

\bibitem[Tre00]{T97}
Luca Trevisan.
\newblock When {Hamming} meets {Euclid}: The approximability of geometric {TSP}
  and {MST}.
\newblock {\em SIAM Journal on Computing}, 30(2):475--485, 2000.

\bibitem[Tre01]{T01}
Luca Trevisan.
\newblock Non-approximability results for optimization problems on bounded
  degree instances.
\newblock In {\em Proceedings of the 33rd ACM Symposium on Theory of
  Computing}, pages 453--461, 2001.

\bibitem[TSSW00]{TSSW96}
L.~Trevisan, G.B. Sorkin, M.~Sudan, and D.P. Williamson.
\newblock Gadgets, approximation, and linear programming.
\newblock {\em SIAM Journal on Computing}, 29(6):2074--2097, 2000.

\bibitem[Vaz01]{V01}
Vijay Vazirani.
\newblock {\em Approximation Algorithms}.
\newblock Springer, 2001.

\bibitem[Vis96]{V96}
S.~Vishwanathan.
\newblock Personal communication to {M.} {H}alldorsson.
\newblock Cited in \cite{Hal98}, 1996.

\bibitem[vL99]{vanLint99}
Jacobus~H. van Lint.
\newblock {\em Introduction to Coding Theory}.
\newblock Springer-Verlag, 1999.

\bibitem[Wig83]{W83}
Avi Wigderson.
\newblock Improving the performance guarantee for approximate graph coloring.
\newblock {\em Journal of the ACM}, 30(4):729--735, 1983.

\end{thebibliography}

\end{document}